FACULDADE DE ENGENHARIA DA UNIVERSIDADE DO PORTO

# Development of Biofeedback Mechanisms in a Procedural Environment Using Biometric Sensors

**Vasco Pereira Torres**

Mestrado Integrado em Engenharia Informática e Computação

Supervisor: Rui Rodrigues (PhD)

Co-Supervisor: Pedro Alves Nogueira (MSc)

# Development of Biofeedback Mechanisms in a Procedural Environment Using Biometric Sensors

**Vasco Pereira Torres**

Mestrado Integrado em Engenharia Informática e Computação

Approved in oral examination by the committee:

Chair: António Augusto de Sousa (PhD)

External Examiner: Maximino Esteves Correia Bessa (PhD)

Supervisor: Rui Pedro Amaral Rodrigues (PhD)

____________________________________________________

July 25, 2013

# Resumo

Ao longo da evolução da indústria de videojogos, os controladores de jogo utilizados nunca tiveram grande variação, apenas modificações dos periféricos mais utilizados como rato, teclado, *joystick*, ou *gamepad*. Só recentemente começaram a surgir novas tecnologias como a Microsoft Kinect, Nintendo Wii ou mesmo a PlayStation Move, que melhoraram de certa forma a interação entre os jogadores e o computador ou consola.

No entanto, estudos concluem que há uma nova tendência no grande tópico da interação homem-computador. *Biofeedback* é a capacidade de um utilizador conseguir controlar certos sinais fisiológicos do seu corpo depois de receber a informação dos mesmos. Essa informação pode ser dada na forma de áudio, imagens ou mesmo um videojogo.

O nosso objetivo nesta dissertação é estudar diferentes abordagens na área de *biofeedback* indireto nos videojogos, de forma a criar uma melhor interação entre homem e computador, e proporcionar uma experiencia mais emocionante e apelativa para o jogador. Para isso focamo-nos na criação de uma *framework* que teste diferentes modelos de *biofeedback* indireto dentro de um determinado jogo, de forma a aferir qual o efeito de cada uma das variações na experiencia de jogo do utilizador. Esta *framework* foi desenvolvida de forma independente ao jogo, com o intuito de poder ser utilizada em futuros estudos.

# Abstract

Before the computer age, games were played in the physical world where players would have to interact with real objects and each other, triggering a series of emotions. Nowadays, the computer games have become one of the most popular forms of entertainment due to their high-level of attraction and accessibility. However, the game industry is always trying to find new ways of making games more interactive and exciting in order to attract new players, and one of the recent trends on the area of human-computer interaction is Biofeedback.

The goal of this dissertation is to study different approaches on the use of indirect biofeedback within videogames, with the purpose of creating a better human-computer interaction, and provide a more appealing and immersive user experience. For this, we focused on the development of a framework capable of testing different indirect biofeedback models within a specified game, in order to assess the effect of each of these variations on the user experience. This framework is game independent, with the intention of being used on further studies.

# Acknowledgments

I would like to thank my supervisors for the continuous support, availability and the provision of great ideas and challenges for this dissertation.

To my family who tried to help me as much as they could even though they did not have much knowledge in the matter. And to all my friends, that discussed with me important topics and helped me with different perspectives.

Vasco Torres

# Contents

# List of Figures

# List of Tables

# Abbreviations

| | |
|---|---|
| AC | Affective Computing |
| AR | Augmented Reality |
| AV | Arousal-Valence |
| BF | Biofeedback |
| BVP | Blood Volume Pulse |
| DDA | Dynamic Difficulty Adjustment |
| ECG | Electrocardiography |
| EDA | Electrodermal Activity |
| EEG | Electroencephalography |
| EMG | Electromyography |
| ERB | Emotional Regulated Biofeedback |
| FEUP | Faculdade de Engenharia da Universidade do Porto |
| GSR | Galvanic Skin Response |
| HR | Heart Rate |
| HRV | Heart Rate Variability |
| N-BF | Non-Biofeedback |
| NV-IBF | Non-Visible Indirect Biofeedback |
| RESP | Respiration |
| UX | User Experience |
| V-IBF | Visible Indirect Biofeedback |

# Chapter 1

# Introduction

Computer games have grown continuously through this last years toward one of the most popular entertainment forms, with a wide diversity of game types and an ample consumer group all around the world. In fact, the computer and console game markets have developed so rapidly that at some point in time it is said that they have surpassed the film industry in terms of total revenues (Barker et al. 2011). Although the game industry continues to expand, it still needs to keep up with the consumer's demands in order to keep growing. From the first games created until nearly 7-years ago, games were associated with traditional input devices, such as keyboard, mouse, joystick or gamepad. Then the era of gestural controls arrived, and the real take-off happened with the introduction of Nintendo Wii, PlayStation Move and Microsoft Kinect in the market, where a new kind of consumers emerged, expanding even more the consumer groups of this industry. Nevertheless new technologies are created every day, even if this kind of games are far from being perfect, there are already new technologies being introduced into games, such as Augmented Reality (AR), with big companies investing on its development, like Google's Ingress[1], a game where the playground is the map of planet earth, and the goal of the game is to aid one's faction to control the majority of the world regions.

[1] http://www.ingress.com/

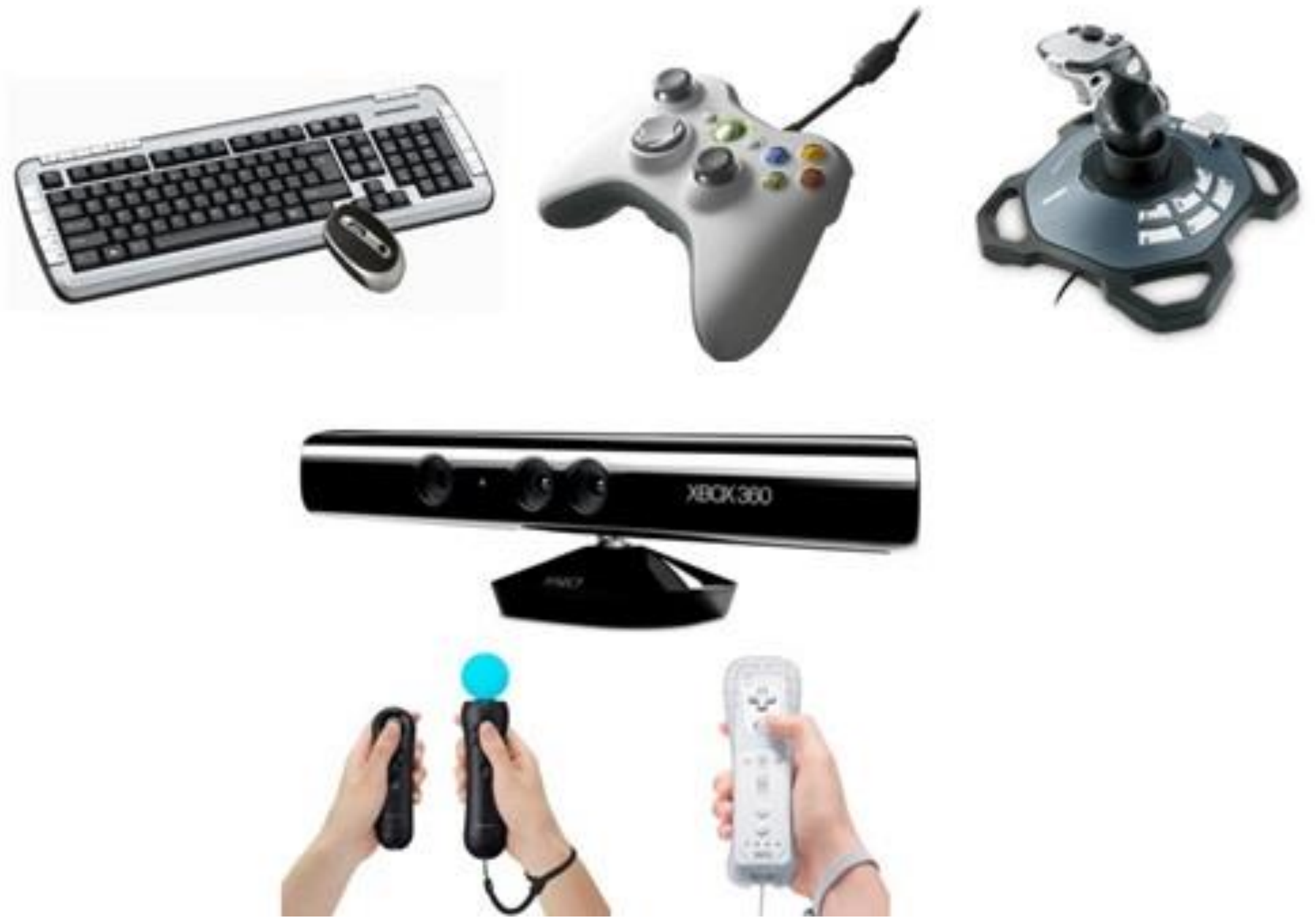

Figure 1: Computer and console controllers.

The development of methodologies for human-computer interaction have been making progress in the last decades in the intersection of computer science, applied sciences of engineering, cognitive sciences of psychology and other fields of study. One of them is Biofeedback (BF), which is the ability of controlling certain physiological or biological functions by receiving information about them. BF can be sub-divided in two main types: direct and indirect. Direct being the use of conscious body reactions such as muscle contraction, and indirect the usage of naturally unconscious physiological functions, such as heartbeat. BF was first introduced in the medical field, but later extended to other scopes, such as computer science and digital gaming. Since the introduction of Affective Computing (AC) in 1995 by Picard (Picard 1995), several studies have been performed on this area, on how to use this new technology to enhance the interaction between human and computer. Concerning the interaction of biofeedback with games, most of these studies rely on direct biofeedback, with the main reason of indirect biofeedback being less perceptible by the players and much harder to design compared to direct biofeedback mechanics (Kuikkaniemi et al. 2010). There are also sensorless approaches (Kotsia, Patras, and Fotopoulos 2012) with the use of devices such as Microsoft Kinect.

Regarding Emotional biofeedback or Affective feedback, the majority of work done has related to stress/relax studies. In these works, it is asked for the player to control his physiological signals in order to achieve a relaxed/stressed state, which in a certain way goes against the principle of affective feedback, since with this approach the player is consciously trying to control his natural unconscious body actions. Further discussion about this matter is presented on State of the Art.

In this thesis we intend to use emotional biofeedback in a more promising way. In previous studies (Bersak et al. 2001) the player is asked to change his emotional state in order to surpass some task or level. This means that, overall the player has to adapt to the game. We will focus on both ways, we want the player to unconsciously adapt to the game becoming more immersed on it, but we also want the game to adapt to the player, in a way that the player can not consciously notice the game play changes, but realizing the improvement on the user experience (UX). For this, we will implement a framework that uses different variants of indirect biofeedback:

- **V-IBF (Visible Indirect Biofeedback) -** Use of the player's physiological functions to adjust certain game mechanics which are perceptible by the player (E.g. avatar speed);
- **NV-IBF (Non-Visible Indirect Biofeedback) -** Use of the player's physiological functions to adjust certain game mechanics which are not perceptible by the player (E.g. map generation, artificial intelligence);
- **ERB (Emotional Regulation Biofeedback) -** Studies the player's emotional reactions to game events throughout the game, and use this information to dynamically trigger/adjust specific game mechanics.

Our goal with this approach is to study how each of these conditions affect the UX, but also the implementation of a framework that can be used on future studies.

This framework will be applied in a survival-horror game called VANISH. A key reason for choosing this game was its procedural mechanics, which was one of our most important requirements. Also, prior studies in this area have been done upon existing games, which made the introduction of game mechanics for biofeedback control, limited. In this project we will not have this problem, since it was developed from scratch and is now being extended for the integration with this thesis.

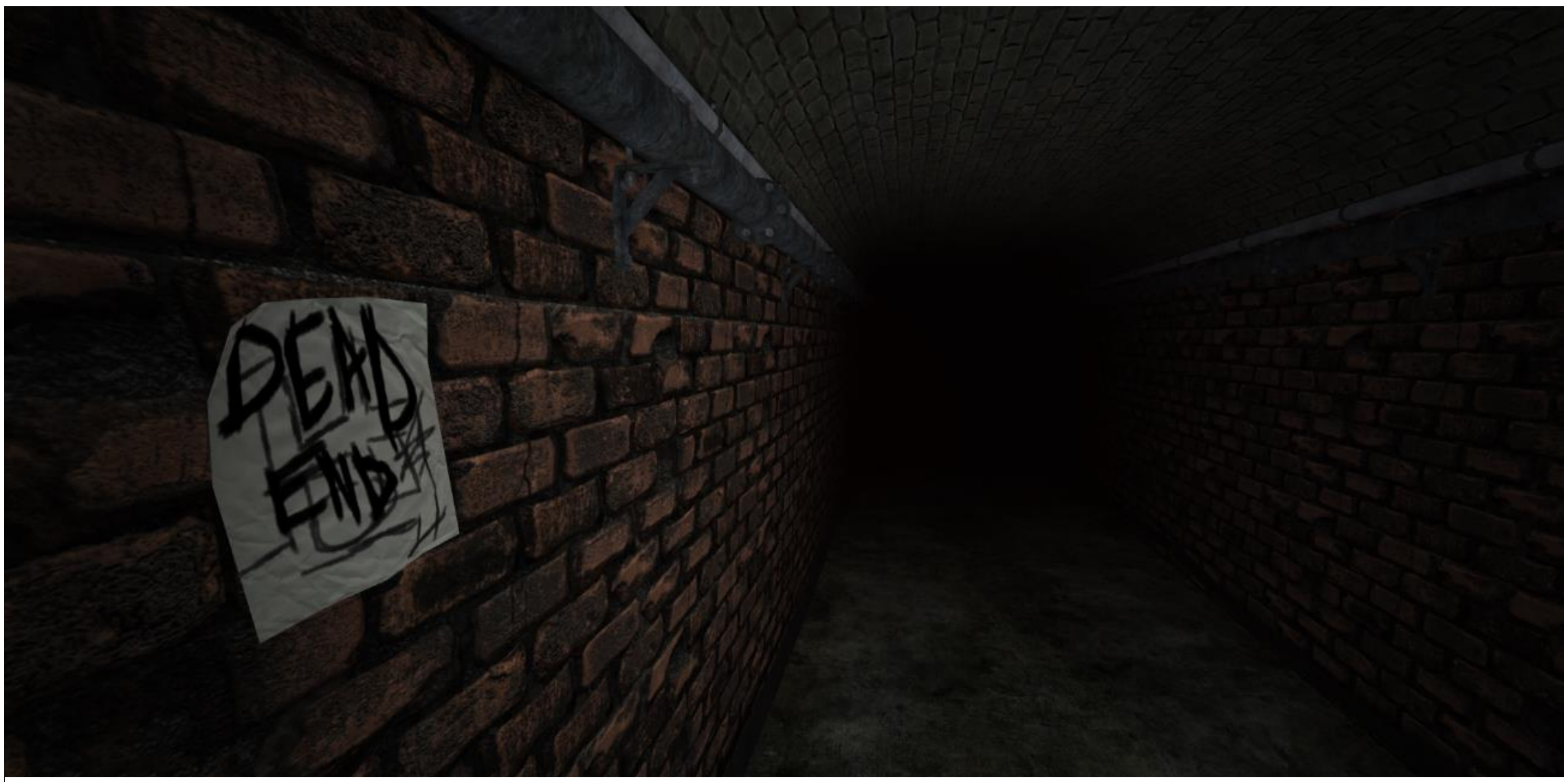

Figure 2: ROAM game play.

Regarding the type of game, most of the works done in this area were based on FPS games such as Half Life 2 (Dekker and Champion 2007), some of them used traditional games like Space Invaders (Lyons et al. 2003) or Tetris (Chanel et al. 2011) and only a few created their own games for the purposes of the study (Giakoumis et al. 2011). Our choice for this type of game was based on the highly emotional variation induced on the player, since the player will face continuous events throughout the game he will most certainly express different affections on each one of them, leading to an easier perception of the player's emotional state and thus giving the opportunity for the game to better match the succeeding game experiences to the player's desires. Also, there is a lack of studies which test the connection between Survival-Horror games and biofeedback.

We strongly believe that we are studying what the future may hold for games, in both technology and user experience, and create a new approach to human-computer interaction. Taking the example of the most unexpected successful game this year, The Walking Dead(2012), during the over exhausted trend of zombie games no one believed on the triumph of this game, except it had one of the best implementations of storyline and game play combined together in a video game. It was acclaimed as an experience that turned and twisted depending on the decisions the player made. This was the key aspect for the title's success: the game play adaptation. However what if, instead of the game asking the player what decisions he wanted to make, it perceived it for itself, through the assessment of the player emotions or physiological reactions and then adapt the game play in order to reflect the player's wishes. We believe this would create a stronger game experience, where the game would change and give the player what he desires. This technology can also be used outside of the game industry. Currently, the prime advertising companies' filter the ads they show based on the user's history, but they could have the actual needs of the users based on their emotional state, giving them the power to customize the ads to a whole new level. We believe we are not very far from this reality.

Our objectives for this thesis are:

I. Implementation of a game independent framework with the purpose of testing different kinds of indirect biofeedback that can be used on further investigations;

II. Adjustment/Development of procedural game mechanics to combine with the biofeedback systems;

III. Development of different models that will link the biofeedback systems to the game mechanics;

IV. Study of the effect on UX of different approaches within indirect biofeedback: V-IBF (Visible Indirect Biofeedback) and NV-IBF (Non-Visible Indirect Biofeedback.

Besides the introduction this document contains more 5 chapters. In State of the Art we discuss the current state of the art in the area. This chapter is divided into 5 sub-sections; section 2.1 introduces the different sensors used for biofeedback; section 2.2 discusses the relevant studies done on the area of direct and indirect biofeedback; section 2.3 examines the various game mechanics present on biofeedback games; section 2.4 provides the reader with information regarding the use of biofeedback for emotional recognition; section 2.5 presents the released industrial applications on the area of biofeedback. Chapter 3 describes the game mechanics from VANISH, and the architecture of our Emotional Engine. The Emotion-Event Triangulation tool is presented on chapter 4. Chapter 5 aims to expose and discuss the obtained results. Finally, Conclusions draws the final remarks on the presented work on chapter 6.

# Chapter 2

# State of the Art

In this section we will present and discuss the different topics that surround the theme of this dissertation. It is divided into 5 sub-sections; section 2.1 provides the reader with information regarding the sensors used for biofeedback systems; section 2.2 discusses the important work done on the area of direct and indirect biofeedback; section 2.3 examines the various game mechanics present on biofeedback games; section 2.4 introduces the reader to the use of biofeedback for emotional recognition; Finally, section 2.5 presents the released industrial applications on the area of biofeedback.

## 2.1 Sensors

A biofeedback system needs to deliver and receive information from the user. In order to receive the data derived from the user's physiological signals, we must use a variety of sensors. Each of these sensors will account for a particular physiological signal.

### 2.1.1 Galvanic Skin Response (GSR) or Electrodermal Activity (EDA)

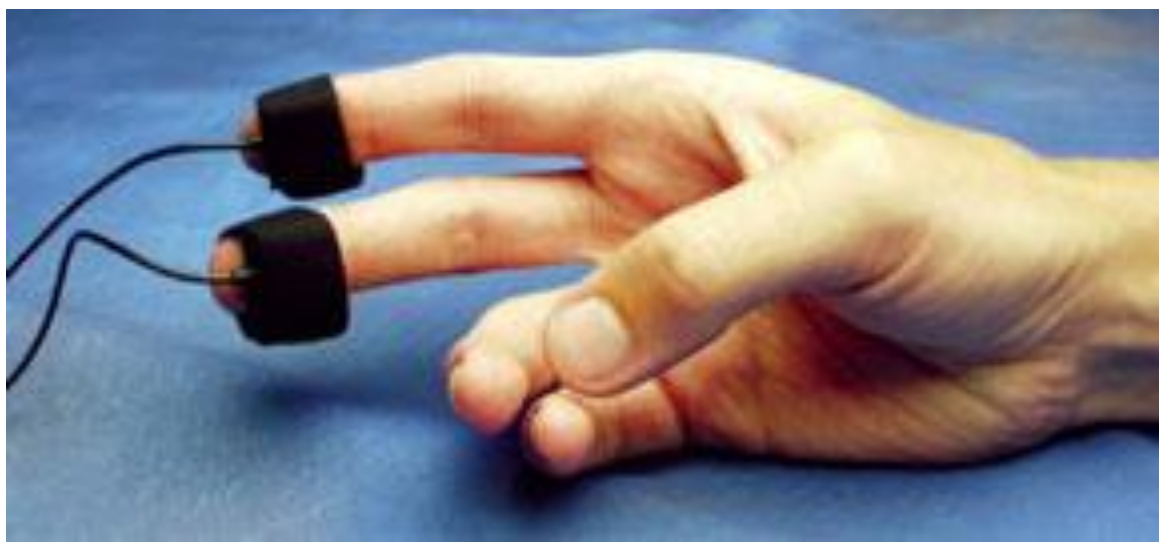

Figure 3: Skin conductance measured through the sweat glands present of finger tips[2].

[2] Adapted from: http://www.biopac.com/Research.asp?Pid=3694&lower=1

This sensor is responsible for measuring the electrical conductance of the skin, which has a variation derived from its moisture level. The sympathetic nervous system controls the sweat glands, thus making the skin conductance a good indicator of physiological arousal.

### 2.1.2 Electromyography (EMG)

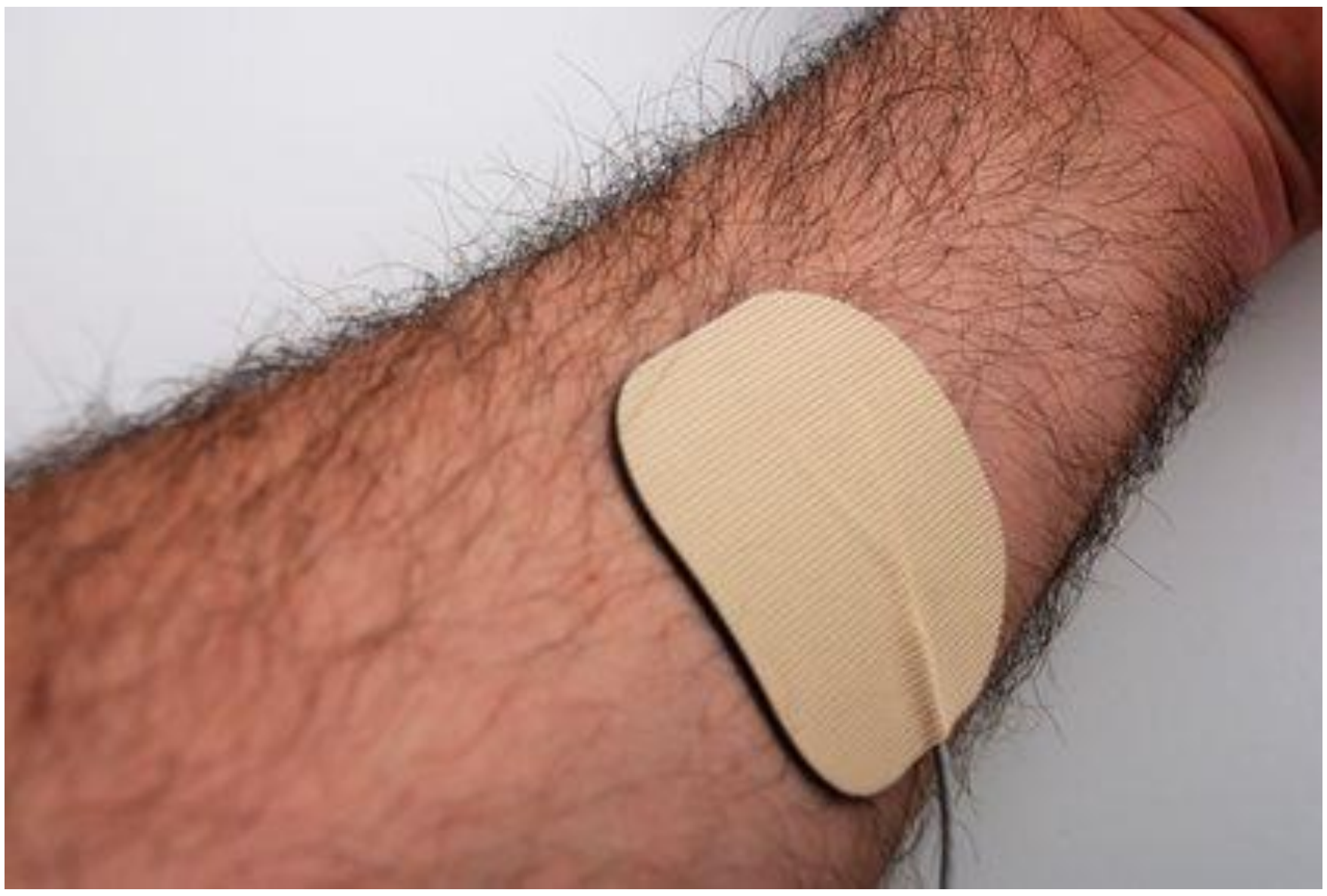

Figure 4: Electromyography on the forearm[3].

Electromyography is used to assess and record the electrical activity produced by skeletal muscles, and it can be used on a major part of the human body muscles. Regarding the practice of this sensor within a biofeedback system, it can be applied on muscles which the user is not always aware of its usage, such as the face muscles, leading to a good association with indirect biofeedback. The other alternative, is the use of this sensor on a muscle that the user has to consciously control, and is better related to direct biofeedback.

[3] Adapted from: http://www.ehow.com/facts_7521460_introduction-surface-electromyography.html

### 2.1.3 Electrocardiography (ECG)

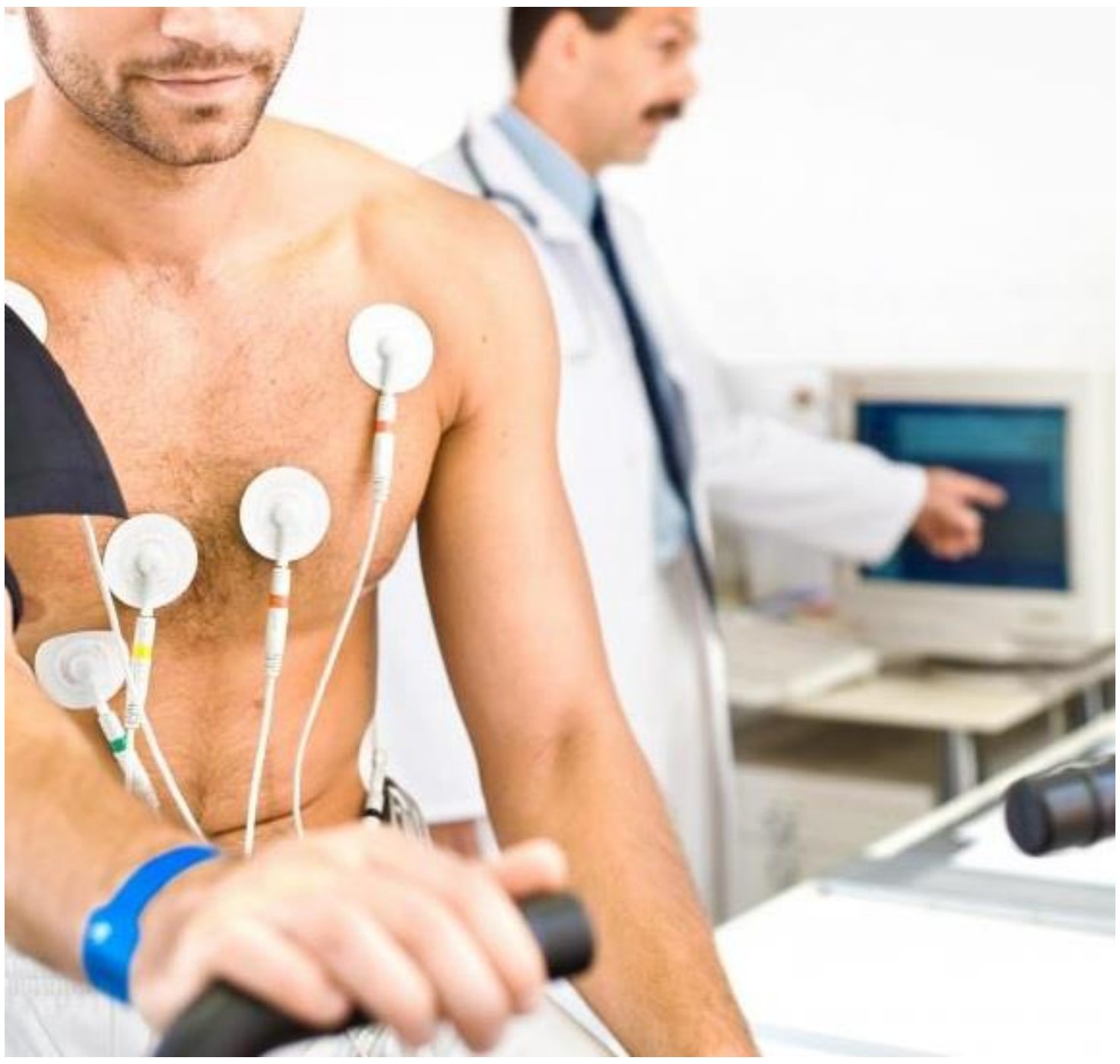

Figure 5: Man performing an electrocardiography[4].

The objective of an electrocardiography is to interpret the electrical activity of the heart over a period of time. This information is detected by attaching electrodes to the outer surface of the skin across the thorax or chest. ECG is mostly used for indirect biofeedback systems, since the user is not typically aware nor in control of its activity.

[4] Adopted from: http://www.wisegeek.com/what-does-an-ekg-technician-do.htm

### 2.1.4 Electroencephalography (EEG)

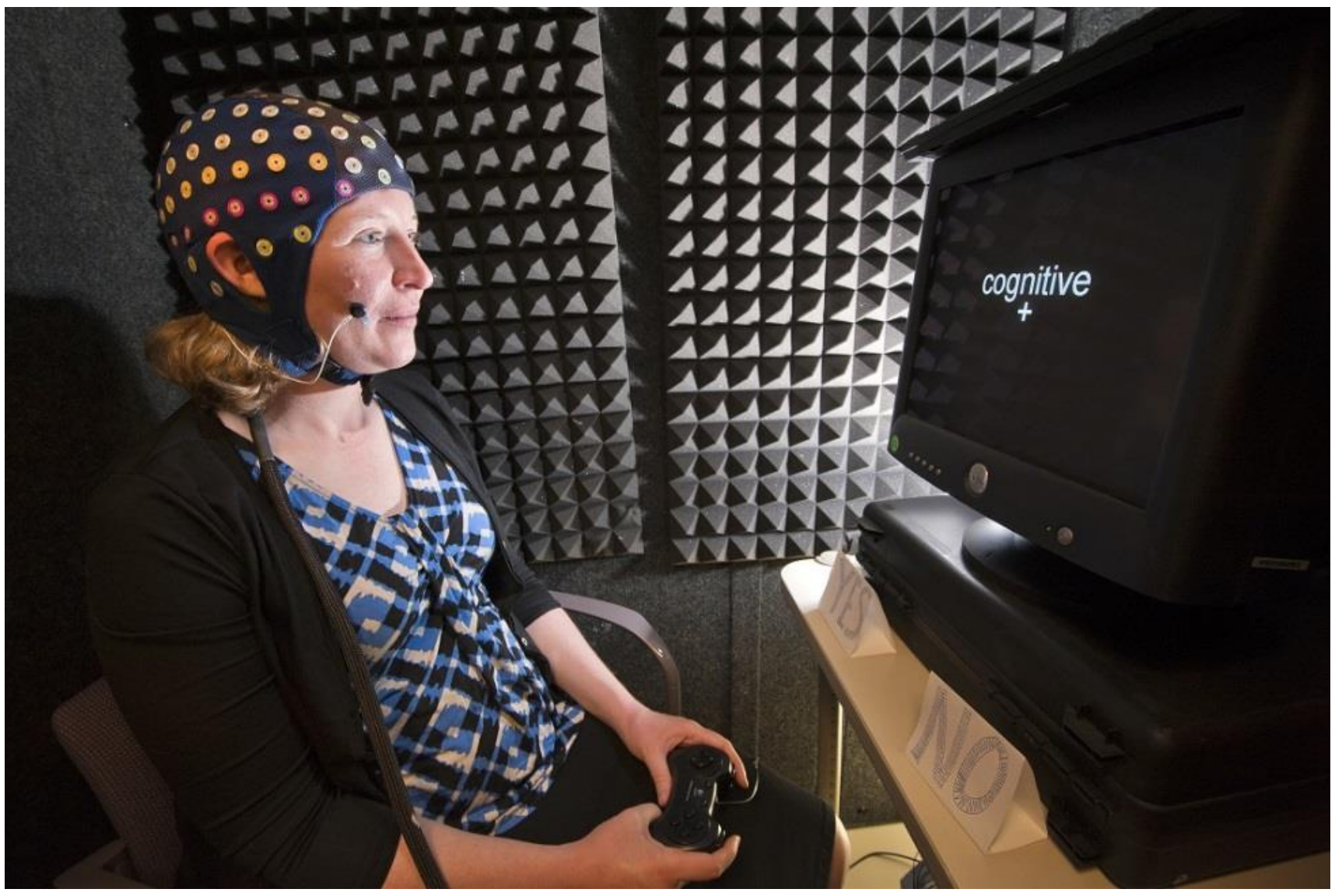

Figure 6: Brain activity monitored during a test[5].

EEG is accountable for evaluating the electrical activity along the scalp, with the purpose of studying the user's brain activity. This sensor tends to be used mostly on indirect biofeedback systems.

[5] Adopted from: https://share.sandia.gov/news/resources/news_releases/brain_study/

### 2.1.5 Respiration (RESP)

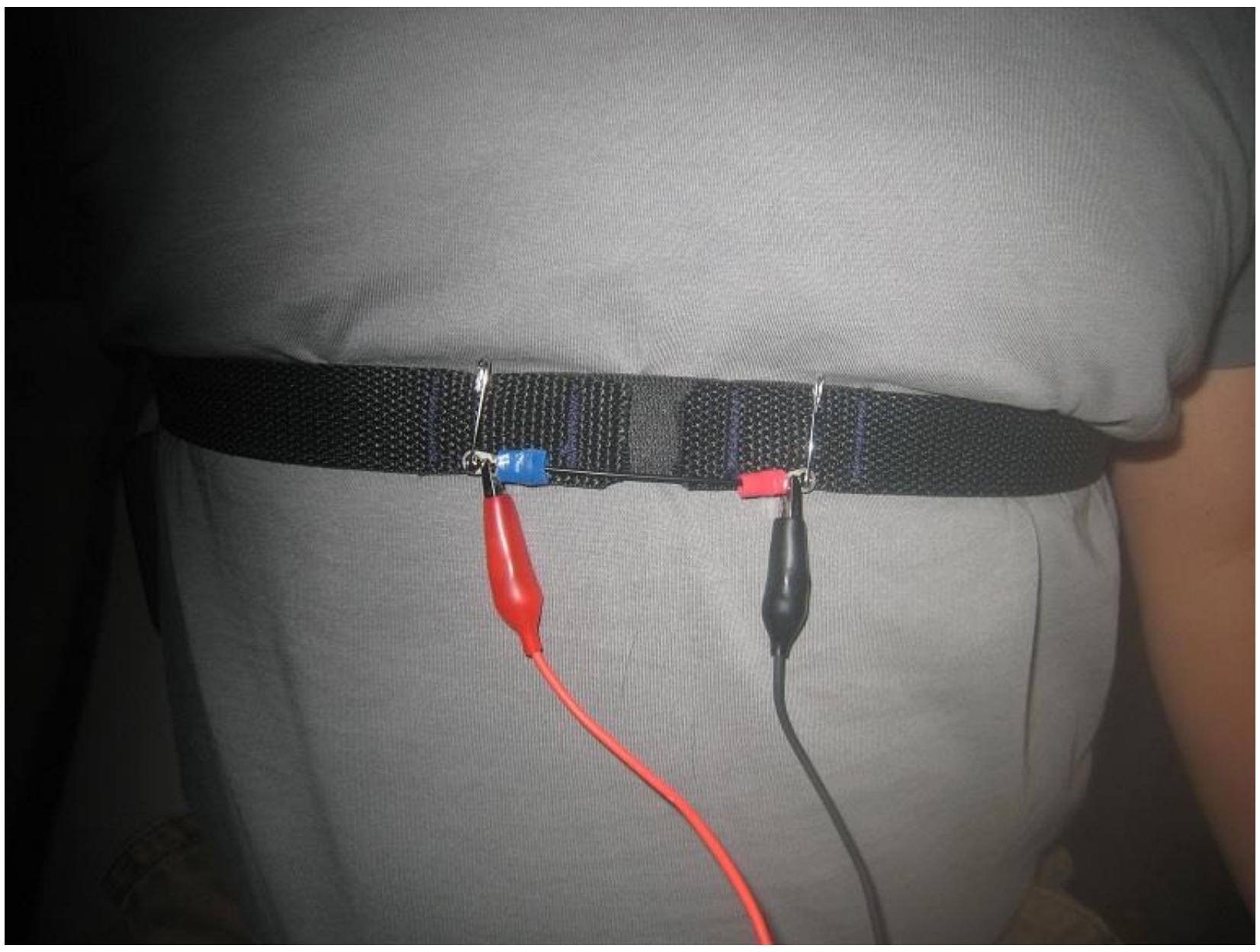

Figure 7: Sensor used to record diaphragmatic and abdominal breathing[6].

The respiration sensor usually consists on a stretch belt that fits around the thoracic or abdominal area, and is used to record the user's breathing. Concerning its applicability on biofeedback systems, it can be used on both direct and indirect biofeedback, due to being a physiological signal which the user is naturally not aware of, however it can easily be controlled after the user becomes conscious of it.

## 2.2 Direct and Indirect Biofeedback

Biofeedback is the ability of self-regulating a person's biological or psychological functions by gaining greater awareness of them with the use of instruments that provide information on those same systems. Biofeedback usually requires the attachment of sensors to the body for the acquisition of biological signals, such as those produced by sweat glands (GSR), heart rhythms (ECG), muscles (EMG), brain activity (EEG) and body temperature. The information pertaining the changes recorded by these sensors is then given to the person by the form of audio, computer graphics or other kind of feedback. Usually on biofeedback systems there are multiple processes, correlating the information from the various sensors and interpreting those values in order to compute a feedback to the user. Depending on the purpose of the system, the algorithms used must be adapted to its specifications, thus requiring a combination of different expertise, making the development of biofeedback systems a complex task.

[6] Adopted from: http://mfleisig.wordpress.com/2010/10/19/diaphragmatic-and-abdominal-breathing/

Biofeedback can be divided in two categories: direct and indirect biofeedback. Direct biofeedback consists on conscious physiological function such as contracting a muscle, and indirect biofeedback corresponds to an unconscious body action such as heart rate or respiration. It is important to note that sometimes the barrier between direct and indirect biofeedback can be easily crossed, for example when the user gains awareness of his respiration and he starts to consciously control his breath. Within Indirect Biofeedback there is Emotional/Affective Biofeedback, where various involuntary sensor channels are interpreted in parallel to infer the user's emotional state.

For instance, direct biofeedback is used on (Lyons et al. 2003) study for the purpose of muscle rehabilitation. There were two kinds of therapy: relaxation and contraction. The game Space Invaders was used for the biofeedback system, where in the contraction mode a fire control command is executed every time the current contraction level of the subject muscle is higher than the threshold set on the calibration (Figure 8). On the other hand if the mode is set to relaxation then the fire control command is triggered when the detected EMG level is below the set threshold. As the study conclusion indicates, the use of a biofeedback system induced an improvement of muscle strength and range of emotion, where the experimental group exceeded their threshold level 136% times more often than the group which did not participated in this study, thus making this an encouraging method for further research.

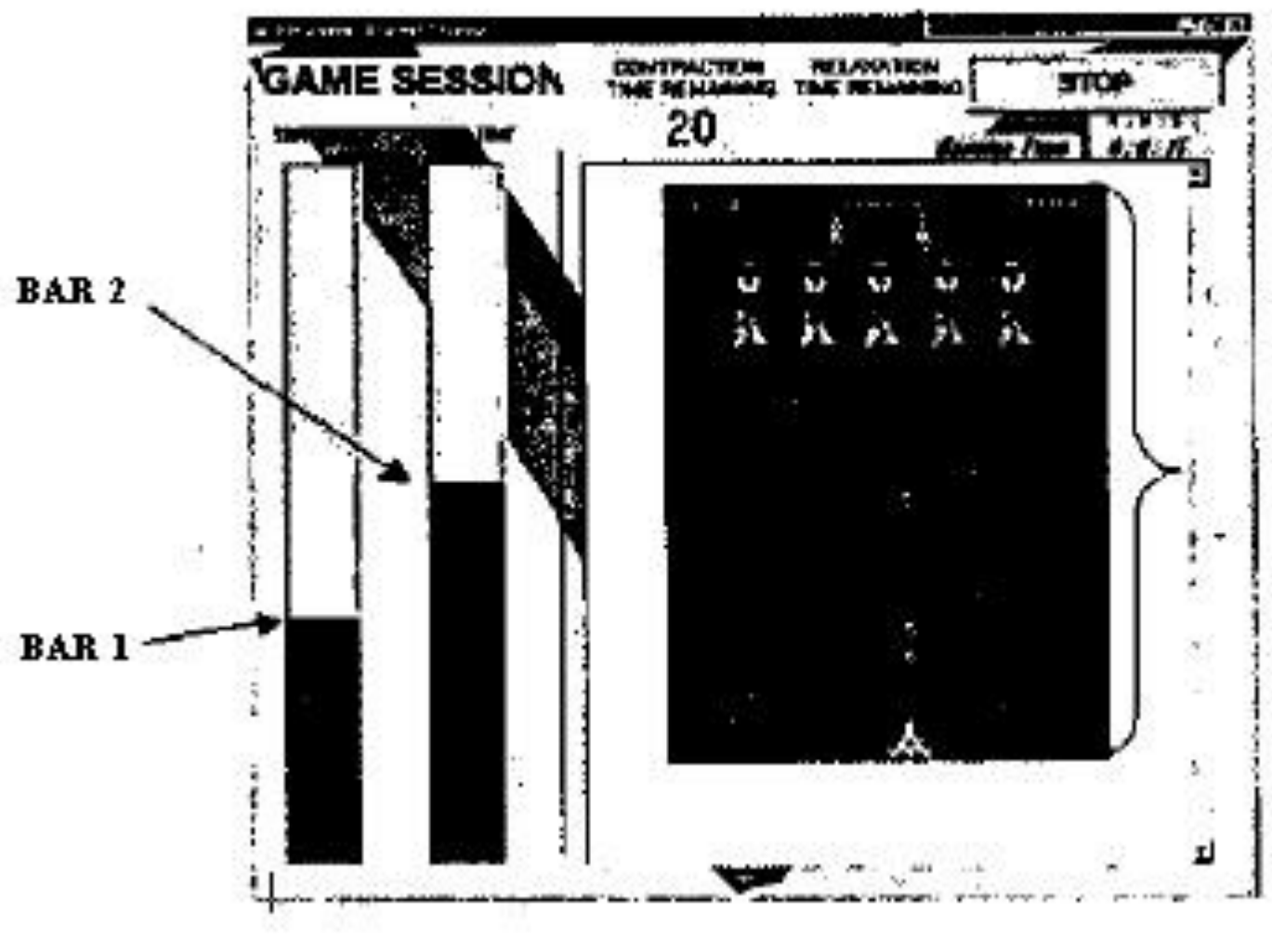

Figure 8: The biofeedback system screen, showing the threshold level (Bar l), the contraction level currently achieved (Bar 2) (Lyons et al. 2003).

Direct and Indirect biofeedback can also be used together as shown by (Nacke et al. 2011). The biofeedback system was developed on a single-player 2D side-scrolling shooter game and used 5 game mechanics to deliver the biofeedback input into game play. Respiration, EMG on the leg and temperature were the physiological sensors used as direct control. The indirectly-controlled sensors consisted on GSR and ECG. These mechanics involved the enemy target size, speed and jump height, the flame length of the flamethrower weapon and in the final scene, the weather (snow falling) and the boss difficulty. These mechanics were controlled using GSR and RESP, ECG and EMG, GSR and RESP, TEMP and ECG sensors, respectively. There was also a special feature which they called Medusa's Gaze, that through the user's eye tracking , showed a circle on the screen corresponding to the place where the user was looking, allowing the player to temporarily freeze its opponents whenever the circle intersected an enemy. The correspondence between these game mechanics and the sensors can be examined bellow in Table 1.

Table 1: Game conditions (Nacke et al. 2011)

| **Mechanic** | **Condition 1** | **Condition 2** |
|---|---|---|
| **Target size** | RESP | GSR |
| **Speed/jump** | ECG | EMG |
| **Weather/boss** | TEMP | ECG |
| **Flamethrower** | GSR | TEMP |
| **Avatar Control** | Gamepad | Gamepad |
| **Medusa's Gaze** | Gaze | Gaze |

The experimental procedure was based on three conditions, 2 of them with physiological control and the other one with ordinary input (gamepad). All participants played all three conditions, which were present in a randomized order. The results were highly satisfactory, when asked whether they preferred to play with or without sensors, 9 out of 10 players preferred to use physiological control. In terms of sensor preference, the most voted sensor was the gaze input (eye tracking), followed by Respiration, and the other sensors had the same number of votes. These results meant that players favored direct over indirect control, due to the visible responsiveness.

In a similar research line, (Kuikkaniemi et al. 2010) compares the use of different kinds of biofeedback: implicit and explicit, where implicit can be associated with the indirect biofeedback definition and explicit with the direct, but further explanation of these ambiguities is discussed on the section 2.2.1. The biofeedback system was implemented on a first-person shooter (FPS) game, and EDA and RESP biofeedback inputs were used for implicit and explicit conditions. When the player became aroused (when the EDA value rose) the character began to shake more, move and shoot faster and with stronger weapon recoil. When the player became more relaxed (evidenced by a decrease of the EDA value) the character moved slower and shot in steadier fashion. In the RESP source, the methods were the same, when the player inhaled the character became slower and steadier. When the player breathed out, the character became faster but his aim was also shakier. The differentiation between the explicit and implicit conditions was set on the experimental procedure. This experimental procedure was divided in two phases; in the first one, players were not aware that the game was being controlled by their physiological signals, making it the implicit condition. The explicit condition was present on the second phase, where the players were told how the system worked and rapidly gained awareness of its features. As the result of this experiment, the RESP source produced no effects on the implicit condition. On the other hand the explicit condition showed many interesting factors, indicating that the players enjoyed playing in this condition led by their great increase in level of immersion.

Both (Nacke et al. 2011) and (Kuikkaniemi et al. 2010) mentioned the difficulty of developing mechanics for indirect biofeedback and the apathy shown by players regarding this type of biofeedback. Both of these studies were conducted on FPS games, which may be one of the causes for the unsatisfactory results.(Nacke et al. 2011) states that:

> "*Indirect physiological control does not have a 1:1 mapping of player action and game reaction and is therefore not equally suited as game input for fast-paced action games… However, this disadvantage could be turned into a strength if indirect physiological control was used to affect slow-changing environmental variables of the game that could allow these sensors to function as a dramatic device.*"

This statement supports our research rationale in 2 critical ways:

I. It supports our claim that indirect biofeedback mechanisms should be used as a method to influence passive aspects of the gameplay experience. In our case, these correspond to the game parameters that regulate the gameplay events that occur and the sanity/stamina mechanics, which the player would not have direct control otherwise.

II. It also supports our choice of game genre, as we will implement these mechanics on a survival-horror game. Besides the deep psychological terror, its generally slower pace with occasional tension and relaxation pikes, mainly characterizes this game genre, thus making indirect control a plausible solution.

(Dekker and Champion 2007) focus only on indirect biofeedback and is also developed on a first-person shooter game, in this case a modification of Half-Life 2. The physiological signals used for the biofeedback system were HRV and GSR. The data coming from the sensors was almost directly used to specific game features, for instance the speed of movement of the avatar was based on the heartbeat multiplier combined with a base level of 200, and like this mechanic others such as audio volume, stealth mode, weapon damage, AI difficulty and a variety of screen effects were also subjective to the information coming from the player's physiological signals. On the evaluation process, subjects played an enhanced biometric level and a standard level. Between the results obtained, there was an unforeseen outcome which showed that audio effects had a significant effect on the participant's biometric information and reactions. They appeared to be more involved in the enhanced version especially when sounds were played.

There are also other approaches within indirect biofeedback, for instance (Toups et al. 2006), that studies how a player physiological signals affect team game play. In their study, the authors use EMG and EDA sensors to manipulate an "activation" variable on PhysiRogue (a modification of the game Rogue Signals). Or even the gender differences regarding cardiovascular reactivity in violent game play which was the scope of (Tafalla 2007), which presented 2 types of game play, one with a violent soundtrack and another one without it. It concluded that men performed twice as well with the soundtrack and the women's performance did not change at all.

One of the most known researches in the area is (Bersak et al. 2001). With the use of GSR to measure the player's stress level, the study implemented a simple game mechanic within a racing game, which they called "Relax-To-Win". As the game title suggests, the player is given the objective of finishing the race first, and for that he needs to relax, considering that the speed of his avatar is controlled by his stress level. If the player becomes stressed, the avatar will move slower, if he calms down, the avatar will increase in movement speed. This mechanic led to a variety of interesting results. First we have the controversial fact, when the player is winning the game he tends to become more stressed, leading to a decrease in movement speed and consequently the possible advance of the enemy. Nevertheless, the same situation happens when the player is losing, as the player gets more frustrated, it leads to an increase in stress level thus falling further behind. Usually when one of the players finish the race first, the loser will become more relaxed and move faster to the finish line. Overall, we consider it as a successful step toward the use of biofeedback within video games.

This last study also brings one of the few known dilemmas in this area. With the exception of studies such as (Bersak et al. 2001), when a game is being adjusted with information provided by the player's physiological signals in an indirect way, it is not generally intended for the player to gain awareness on how the mechanics work, as this allows him control over the system. If this happens, the player can start to cheat his physiological actions and mislead the game. A possible solution for this problem is the development of game mechanics with no direct or apparent connection with the information supplied by the sensors. Another approach may be to delay the adjustment of game features, albeit in a relevant time period, otherwise the timing might be missed and the player's engagement lost.

## 2.3 Game Mechanics

### 2.3.1 Biofeedback and Affective Gaming

There can easily be misinterpretations when analyzing the different studies done on the biofeedback and affective feedback field, as the employed terminologies are not always congruent. As explained in section 2.1, the definitions and boundaries between direct and indirect biofeedback are relatively clear. However, there are studies that refer to implicit and explicit biofeedback (Kuikkaniemi et al. 2010). In this case, despite the initial resemblances, implicit biofeedback does not explicitly correlate with either direct or indirect biofeedback but rather with a third type of feedback: affective feedback. On the other hand explicit biofeedback refers to any kind of biofeedback. Though, what characteristics distinguish biofeedback from affective feedback and what is it exactly?

As stated by (Bersak et al. 2001) affective feedback refers to "*in essence .. that the computer is an active intelligent participant in the biofeedback loop*", where both player and game are affected by the actions of the other. The main difference between affective feedback and biofeedback is conscious or unconscious control of the player's physiological reactions. On affective feedback the player should not even be aware that their physiological state is being measured during game play in order to prevent the conscious control of that state, on biofeedback the player may explicitly control his physiological responses in order to control the different features inside the game.

(Gilleade, Dix, and Allanson 2005) also points the fine line between these two kinds of feedback, that most of the time tends to become fuzzy. Taking for example (Bersak et al. 2001, Dekker and Champion 2007); both these studies use indirect physiological signals, such as GSR, to create indirect biofeedback mechanisms and ask the player to intentionally control their excitement level in order to gain some advantage in the game. What seemed to initially have started as an affective biofeedback mechanism-based game has become an indirect biofeedback-based one due to the awareness of the player in controlling those aspects of the game.

It is important to clarify that an affective gaming experience is the one where it is possible to maintain an affective feedback loop without the player's consciousness of his contribution to that loop. Once that loop is interrupted by a conscious action of the player towards it, the essence of the affective loop is lost and it becomes a form of biofeedback.

### 2.3.2 Dynamic Difficulty Adjustment (DDA)

In recent years there have been several studies investigating the feasibility of dynamic difficulty adjustment (DDA) mechanisms for computer games. The aim of these DDA mechanisms is to adapt the gaming experience to the player's unique performance indexes without human intervention. Despite the fact that most of these works center their attention on some kind of level, character or gameplay performance metric for game play adaptation (e.g. how much hit points the player lost in the last enemy encounter), we believe that the player's affection state can also pose an interesting performance metric in terms of the player's gameplay experience and provide a useful indicator for a DDA mechanism.

(Yannakakis and Hallam 2009) investigate the creation and consequent implementation of a DDA model on the game Bug Smasher. This model performs frequent adjustments throughout the game to regulate specific game parameters to augment the entertainment value of the player. The performance of this adaptation mechanism is then estimated using a game survey experiment. Since the different models studied have a wide complexity of parameters that reside outside the scope of this thesis, we refer the reader to (Yannakakis and Hallam 2009). During the game, different "bugs" appear sequentially on the game surface and disappear after a short period of time. The bug's location is picked randomly according to a predefined level of spatial diversity, measured by the entropy of the bug-visited tiles. On the adaptive variant of the game, the position and the speed of the bugs ascending and descending action was controlled by the model in order to provide a greater experience to the player. In other words, the game adapts to the specific player by attempting to maximize the generated entertainment value. The results of this study concluded a preference of 76% for the adaptive version of the game.

Another example of affective biofeedback is the work presented by (Changchun et al. 2009), whom developed a mechanism that analyzes the player's physiological signals to infer his or her probable anxiety level. Based on these anxiety levels, a pre-defined rule set was used to choose the DDA policy, which was then used to automatically adjust the game difficulty level in real time. The peripheral physiological signals that were measured through biofeedback sensors were: features of cardiovascular activity, including interbeat interval, relative pulse volume, pulse transit time, heart sound, and preejection period; electrodermal activity (tonic and phasic response from skin conductance) and EMG activity. The selection of these signals was based on the liable chance to demonstrate variability as a function of the player's emotional states. An experimental study was conducted to evaluate the effects of the affect-based DDA on game play by comparing it with a performance-based DDA. The games played consisted of solving an anagram and playing Pong. On the affect-based dynamic difficulty adjustment system they classify anxiety in three levels- low, medium, and high. Figure 9 shows the state-flow model of this method, where it can be seen that low anxiety results on the increase of difficulty level, medium anxiety causes no change, and high anxiety outcomes in a decrease of difficulty level.

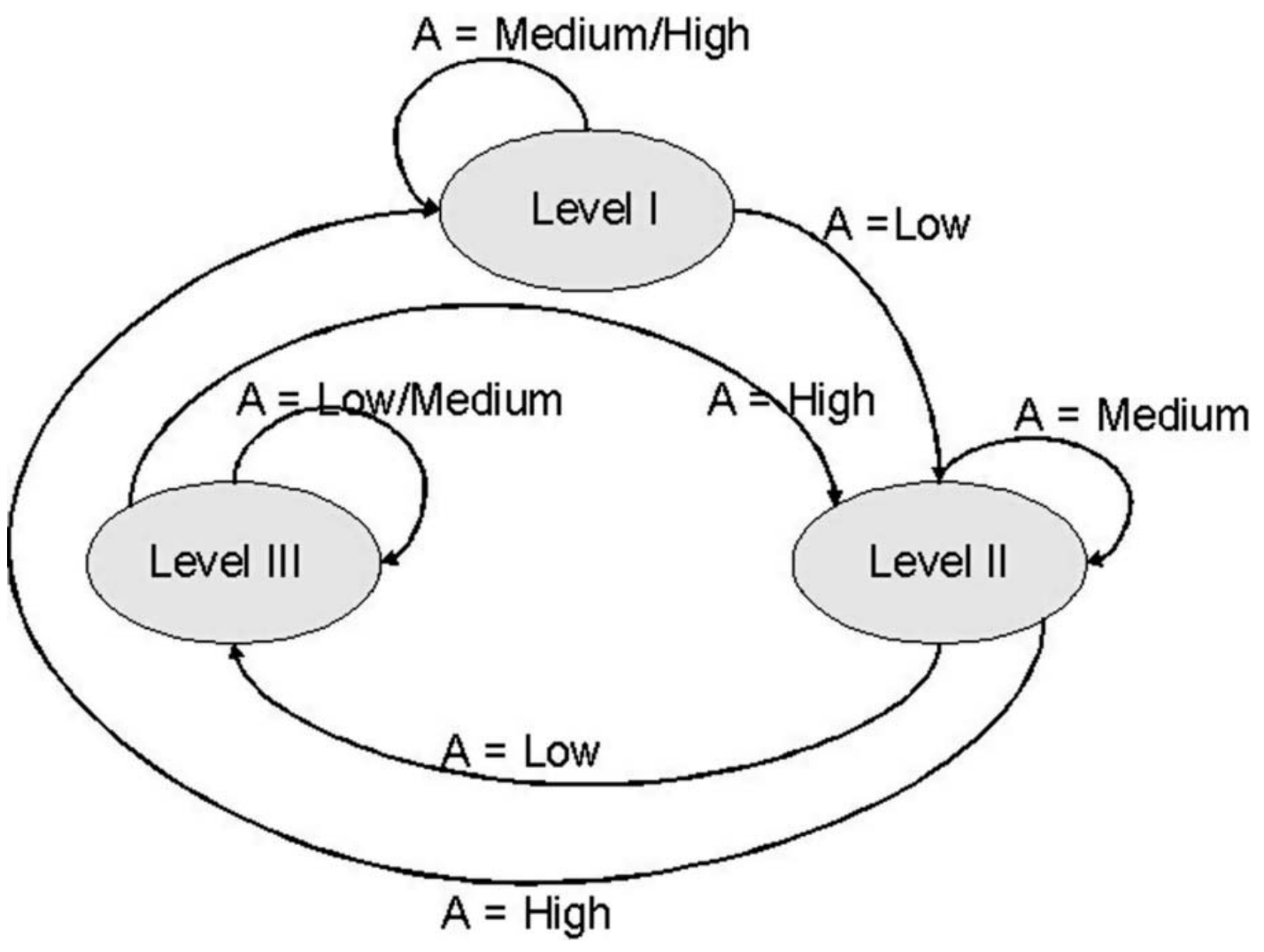

Figure 9: State-flow model of the affect-based dynamic difficulty adjustment system (Changchun et al. 2009).

Based on the test results, the authors concluded that physiology-based affective modeling provided a promising methodology for objectively quantifying player’s emotions when interacting with computer games, but also mentioned that further exploration in this direction is needed. Similarly (Guillaume, Konstantina, and Thierry 2012) studied the possibility of automatic emotion assessment through several peripheral signals in order to provide emotional feedback to the computer game Tetris, and adapt the game difficulty level, in this case the speed of the falling blocks. The considered emotional states, anxiety and boredom are then used to adjust the Tetris level to the player’s skills.

### 2.3.3 Procedural Mechanics

Early computer games had serious constraints regarding memory usage. This implied that a good part of game content had to be generated "*on the fly*"; there was just not enough space to store large amounts of data. Several techniques were used for this purpose, such as Pseudorandom number generators (Barker et al. 2011) used to create very large game worlds that appeared premade. A distinguished example is Rescue on Fractalus, which used fractals to generate the content of an alien planet. These techniques continued to be used, and are present on a lot of popular games in a vast diversity of game genders, in particular in games such as: Grand Theft Auto, Gran Turismo, The Elder Scrolls, Diablo, Left 4 Dead and more recently Minecraft.

Our idea is to intersect these mechanics with biofeedback. Since the biofeedback system consists on a continuous loop that is cycled throughout the game it seems like a pleasant solution to have this cycle feed information to the various procedural mechanics in order to obtain a more efficient and dynamic system.

As seen in (Yannakakis and Togelius 2011), there are already a few studies approaching this idea, analyzing different player experience models (Figure 10) and how affective gaming can be introduced in these systems.

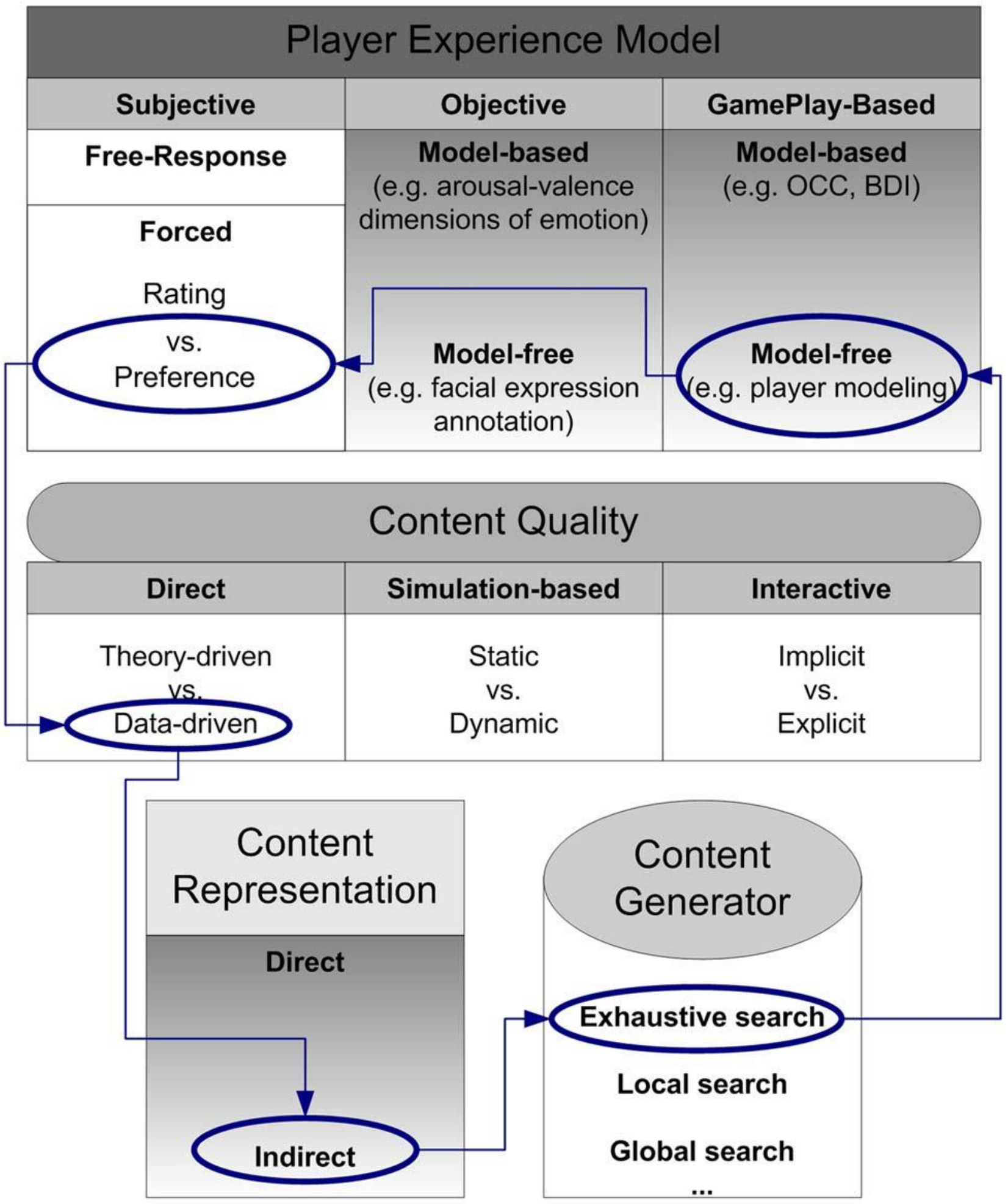

Figure 10: The Experience-Driven Procedural Content Generation (EDPCG) framework in detail (Yannakakis and Togelius 2011).

## 2.4 Emotional Recognition

Regarding the different approaches when assessing the player's emotions, there are two main perspectives. The first perspective describes that some emotions are present in humans from the early days of their birth, in a sense that those emotions can be adapted later on to a specific value, without crossing a particular threshold - this is the discrete emotion theory also named Ekman's basic emotions. The second perspective is the dimensional theory that categorizes all kinds of emotions in a 2-dimensional space, postulating that every emotion has two aspects: a cognitive (Valence) and a physiological (Arousal) component.

Anger, disgust, fear, happiness, sadness, and surprise are Ekman's basic emotions. On his continuous studies he revealed the possibility of adding new emotions to the list using a set of individual characteristics to distinguish the different emotions (Ekman, Dalgleish, and Power 1999).

The dimensional theory crossed diverse approaches, starting with the possibility of describing emotions with 3 dimensions: "pleasantness–unpleasantness", "attention–rejection" and "level of activation" (Schlosberg 1954), and reaching a close consensus with the use of 2 dimensions in the current models. Nevertheless, nearly all dimensional models incorporate valence and arousal or intensity dimensions.

In this thesis we will focus on the dimensional theory for emotions, thus a shortly introduction to this system is needed. Valence describes the pleasantness or hedonic value and Arousal the physical activation. An emotion such as joy and exhilaration is modeled as high arousal and high valence; on the other hand stress would be high arousal but low valence. In order to map these two components to the player's physiological actions we will resort to a variety of biometric sensors.

### 2.4.1 Sensors Used for Emotion Recognition

In this section we will briefly introduce what sensors were used for either arousal or valence on previous works.

As denoted by (Stickel et al. 2009) the valence component is difficult to measure, as it consists of cognitions. The most accessible way of determining valence is with the right questions and questionnaires, but there are, however, also approaches to assess and calculate the valence from physiological data. The arousal element can be easily measured by physiological methods.

(Nacke, Grimshaw, and Lindley 2010) studies the measurement of the user experience regarding sound on a first person shooter, and uses EMG to account for valence and EDA for arousal. The study conclusion features the insignificant effect of neither sound nor music, nor the interaction of sound and music on both EDA and EMG. (Aggag and Revett 2011) uses GSR to account for both arousal and valence with the aid of questionnaires and confirmed the successful use of GSR and its easiness to collect data compared to other sensors. EEG can also be associated with arousal as seen by (Chanel et al. 2006) and it can be an interesting application with other physiological signals. Likewise, the different HR variants can be used for arousal, being one of the most used physiological signals for this component. Some of the studies that use this approach are (Luay and Revett 2011) and (Drachen et al. 2010). There are also studies like (Gu et al. 2010) which concentrates on the performance of a vast number of sensors (ECG, BVP, RESP, EMG, GSR) individually or the effectiveness of multiple sensors correlated with each other (Iancovici, Osorio, and Rosario Jr 2011). A significant part of the mentioned works stated that further investigation must be done for the correlation between different physiological signals and their use for either valence or arousal.

### 2.4.2 Applications and Models

The studies on the assessment of the player's emotions correlated with games have been done with different approaches, some focus on the evaluation of the player's emotions on existing games without changing them, with the intention of studying the applicability of certain sensors (Drachen et al. 2010) or to develop new models for emotion recognition (Mandryk and Atkins 2007). Others create new mechanics or adapt specific aspects on released games to measure the variances on the player's physiological reactions, with a few of them comparing the two versions of the game (Nacke, Grimshaw, and Lindley 2010, Aggag and Revett 2011). Lastly, several of them created new games for the purpose of the study (Giakoumis et al. 2011, Luay and Revett 2011, Groenegress, Spanlang, and Slater 2010).

Concerning the generation of the user's emotional status, a series of models have been studied. Starting with the approach of (Mandryk and Atkins 2007), that modeled the data resultant from the user's physiological signals in two parts, using a fuzzy logic approach. Initially, they computed the arousal and valence values from the normalized physiological signals (GSR, HR, and EMG), and afterwards used this values to produce emotion values for boredom, challenge, excitement, frustration and fun. The fuzzy system used to translate the physiological data to the Arousal-Valence (AV) Space can be seen in Figure 11.

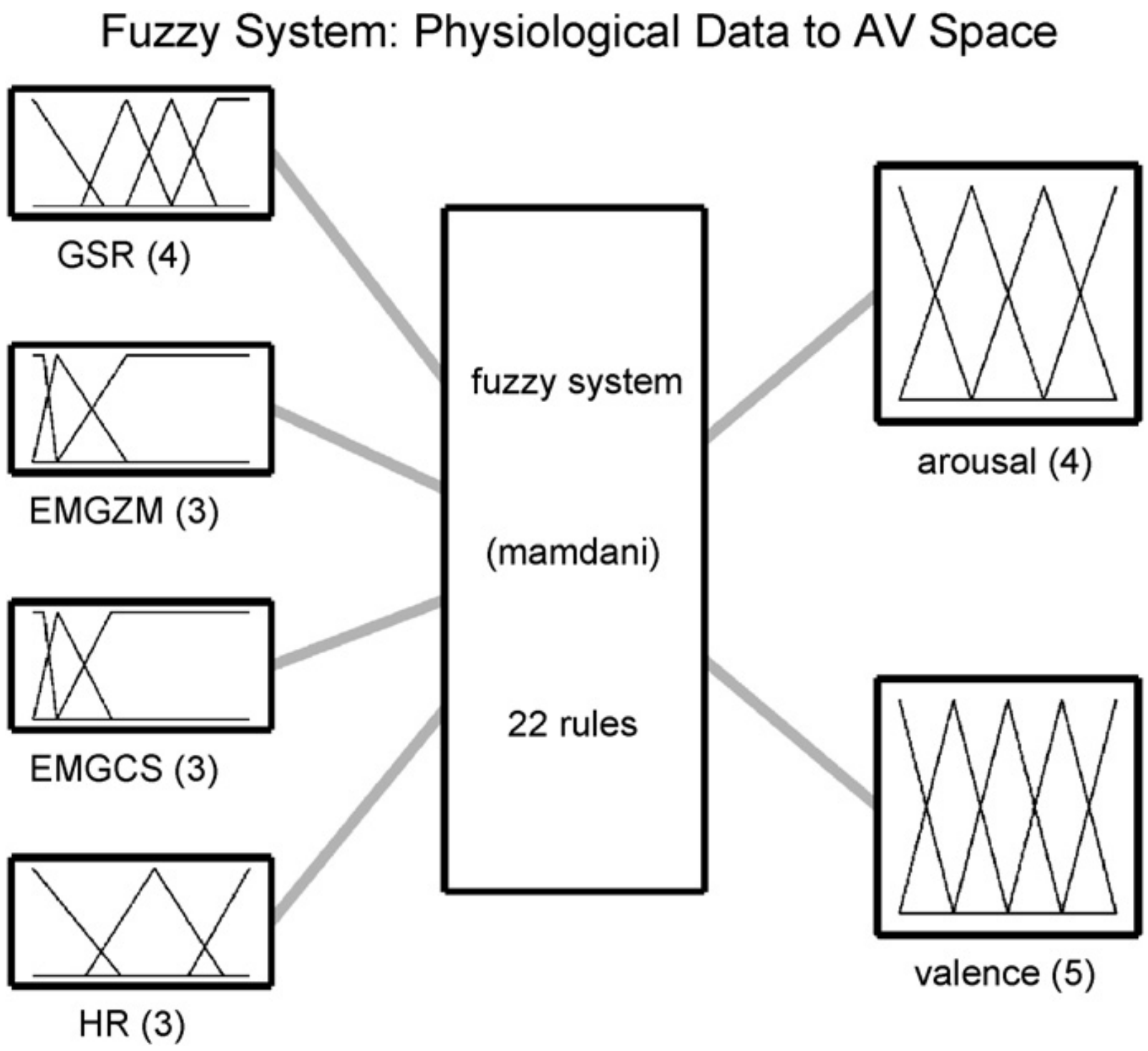

Figure 11: Modeling arousal and valence from physiological data (Mandryk and Atkins 2007).

Another method was used by (Drachen et al. 2010), that consisted in 3 levels of 1 independent variable (games Prey, Doom 3 and Bioshock) and 3 dependent variables to quantify the player's emotions. These variables were: HR and EDA from the player's physiological signals; and the In-Game Experience Questionnaire (iGEQ), which is a short self-report scale for exploration of player experience during playing a digital game. To the values from HR and EDA was applied a simple normalization, where the average value for the feature for each 5-minute segment was divided with the average value from all three baselines recorded per participant. By normalizing the HR and EDA data with the average HR/EDA, they isolated the game-related effects. Finally, the Pearson's correlation coefficient was calculated between the seven dimensions of the iGEQ questionnaire and the normalized physiological data across the three games. The values obtained are displayed on Table 2.

Table 2: Pearson’s correlation coefficient between iGEQ dimensions and physiological measures (Drachen et al. 2010)

| **Physiological measures** | **Competence** | **Immersion** | **Flow** | **Tension** | **Challenge** | **Negative affect** | **Positive affect** |
|---|---|---|---|---|---|---|---|
| **HR** | -0.36 | -0.43 | -0.25 | 0.37 | -0.31 | 0.24 | -0.42 |
| **EDA** | -0.08 | -0.23 | -0.24 | 0.02 | -0.18 | 0.38 | -0.20 |

## 2.5 Industrial Applications

As popular media has shown, correlating biofeedback with games (or virtual interactivity) appears to be something the mainstream culture longs to see. Films like The Matrix or Surrogates, discuss the implication of being physically connected to virtual reality system or mechanical body that responds to the participant's physical state.

The current state of innovation suggests that we are still far away from the reality presented on these films. Regardless, numerous mainstream game developers and publishers, such as Valve software, have studied the integration of biofeedback sensors into some of their existing products, with the primarily reason of gathering player physiological output as part of usability testing. It is only implemented for research, thus making it not available for the end user. Furthermore, Valve also declared "*We're frustrated by the lack of innovation in the computer hardware space though, so we're jumping in", "Even basic input, the keyboard and mouse, haven't really changed in any meaningful way over the years. There's a real void in the marketplace, and opportunities to create compelling user experiences are being overlooked.*" This may lead to new advances in this field.

There are other examples of developed biofeedback-based games, unfortunately not all of them reached the consumer market:

- Atari Mindlink (1984), a unreleased video game controller consisting on a headband whose sensors are supposed to pick up facial movements and muscle actions, in order to control the movements of the paddle and use it as input instead of the ordinary gamepad/joystick. There were two games in development for this controller, Bionic Breakthrough, a "bounce the ball into a brick" game, and Mind Maze which is a game played somewhat like those old mind reading experiments where a scientist would hold up a card and ask a person to tell him what was on the other side. Depending on the game selected two to four cards appear on the screen each round and the player must try and guess the "correct" card by highlighting it using the controller.
- Oshiete Your Heart (1997), a Japanese dating game where the heart rate and sweat level of a player is measured in order to influence the outcome of a date.
- Journey to Wild Divine (2001), a biofeedback game that requires the player to adjust his heart rate variability and skin conductance level to navigate through a series of adventures in a video-game type interface.

- Mindball (2005), a two-person game that consists on the assessment of the player's brain waves in order to compete for the control of a ball's movement across a table.
- MindWave (2010), a brain-wave headset from the manufacturer Neurosky. A product aimed for both education and entertainment. There are already plenty of games and apps available to be played using the Neurosky headsets, most of them being mental training programs. Even if some of the apps are more game focused, they still lack the expected gameplay experience as traditional games.

More recently Nintendo has announced the Wii Vitality sensor, but also stated the difficult time performing consistently across a variety of situations making it less likely to be released any time soon. Similarly, Ubisoft announced in 2011 their pulse oximeter sensor, called "Innergy".

# Chapter 3

# Biofeedback Framework

In this section we will describe the developed biofeedback framework, all the while presenting the reader with a detailed explanation of each one of its constituent components. To develop certain elements of this framework there was a need for an adaptable test-bed videogame that could provide us with the necessary control over its gameplay mechanics and event system. Given the readiness and procedural content generation, we chose to adopt our own indie game VANISH – which was already in development in a personal pet project. This chapter is divided in two main sub-sections: In sub-section 3.1 we describe VANISH and its gameplay mechanics, since they will be used to define our biofeedback conditions. On the sub-section 3.2 we will describe how the Emotion-Engine ($E^2$) (Nogueira et al.) architecture was adapted for this particular study on IBF mechanisms by providing the reader with detailed information on each of its components.

## 3.1 VANISH

VANISH is survival-horror game with the key aspect that the decision on which events and map sections are to be created is driven by a parameterized stochastic process, and the creation aspect of the game is procedural, following a group of rules with the purpose of delivering a better UX. This means every new gameplay session is a different experience for the player.

It would be interesting to know how the game providing different experiences impacts the study. On one hand it is necessary for certain events/maps sections to be adapted by us. On the other hand, it might make it more difficult for players to spot differences between conditions, but is a necessary risk that must be taken given the degree of experience alteration freedom we want to achieve. This implies great care must be taken in developing the gameplay modification mechanisms in a way that they are significantly noticeable by the players, but do not break the game's balance by making it either too hard or easy to win.

### 3.1.1 Game Progression

When the game starts, the character will wake up facing a door that is closed and cannot be opened by the inside. He is then left with no other option but to start exploring the surrounding areas. As the character starts roaming, he will find himself on a series of tunnels, all of them with a similar shape. The tunnel's visual aspect resembles that of sewers, but there is no water on the floor, just pipes travelling through the walls and lights sporadically attached to the walls here and there. Soon enough the character perceives that he is not the only living creature roaming this sewer maze. After a certain (semi-random) period of time, odd events start to take place, with light bulbs bursting with no apparent reason and pipes breaking or falling to the ground. The game's atmosphere continues to progressively deepen as the player starts finding chalk markings and cryptic papers on the walls that do not make sense. As the character roams through the tunnels, his sanity finally shatters when he apparently hallucinates, seeing two glowing eyes in the distance (Figure 12).

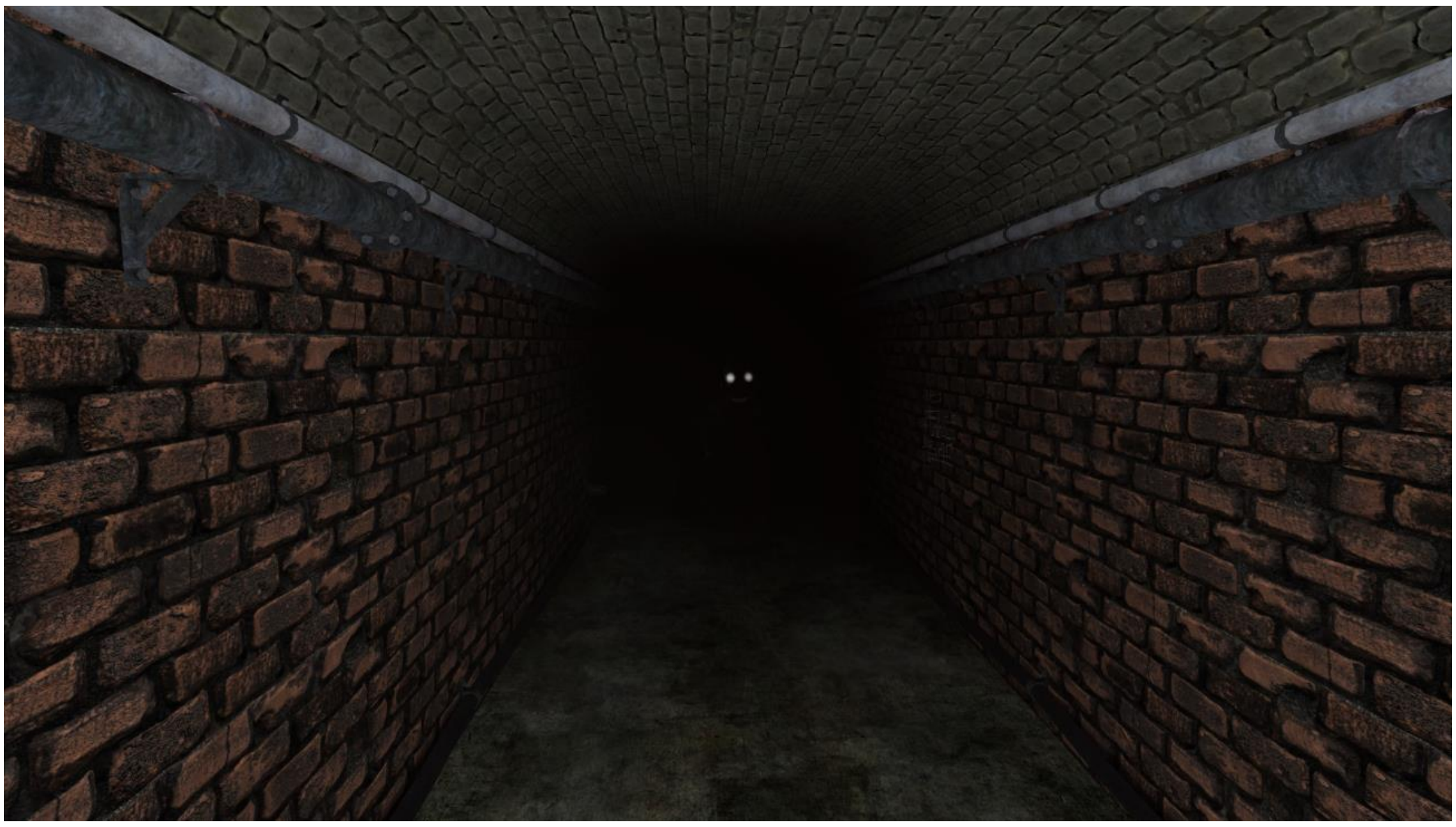

Figure 12: Roam gameplay, creature eyes seen in the distance.

After walking for some minutes, the player will find wide, open area rooms that break the game's architectural motif so far and aid the process of imaginative immersion. On the floor there of these rooms there may be a folder that provides narrative information. These notes shed some light into the previously aforementioned events and aid the character in understanding what is really happening. If the character is eager to survive, he will try to make his way back to where he woke up and attempt to save his life by trading the acquired information.

### 3.1.2 Winning/Losing Conditions

In order to complete the game, the player has 3 objectives. The first two objectives rely on acquiring folders that are present on Key Rooms. These rooms have a series of mechanics that will make them considerably hard to find (see 3.1.3). To pick up a folder, the player must look at it and press the Left Mouse Button (LMB). Once the player does so, a tooltip text will be shown on the top part of the screen, indicating how many folders have been picked so far.

Once the player has both folders, he must attempt to reach the room where he/she first spawned on the game world. Whenever the player reaches that room with both folders he/she wins, and the game will switch to the winning menu.

The losing condition is fairly simple; if the creature catches the player, a death animation plays and the game switches to the losing screen. In this screen the game camera remains in the first-person perspective and the creature is dragging the player's helpless avatar through the tunnels towards its nest.

### 3.1.3 Procedural Level Generation

The game's level is continuously generated during gameplay. This means that there is no unique map, and every gameplay will have a unique – possibly idiosyncratic – map design.

There is a set of 10 map blocks that can spawn on the game world (Figure 13). Each one of them with certain characteristics, and can be divided in the following groups:

- **Normal Tunnels** - This group consist of straight, corner, 3-way and 4-way blocks. These are the blocks that spawn more frequently and they have no special characteristic or purpose other than to provide spatial interaction and host gameplay event mechanisms;
- **Key Rooms** - These blocks contain a folder, which constitute one of the objectives of this game. They are large open areas with 4 entrance/exit points. There are two of these rooms, one for each folder;
- **Exit Room** - This is the block where the player starts, and where he has to go when he picks up both folders, it has only one entrance/exit point;

- **Evasion tunnel** - This is a special block that has the shape of a straight block, but with a small modification: it has an "*evasion tunnel*". This evasion tunnel is a hole in the wall with the height of a child that the player can use to hide from the creature. To enter into this evasion tunnel the player must assume a crouching stance. This requires a certain degree of agility (i.e. control proficiency), which increases the sense of tension – and competence, if properly executed –, in a chase scenario.

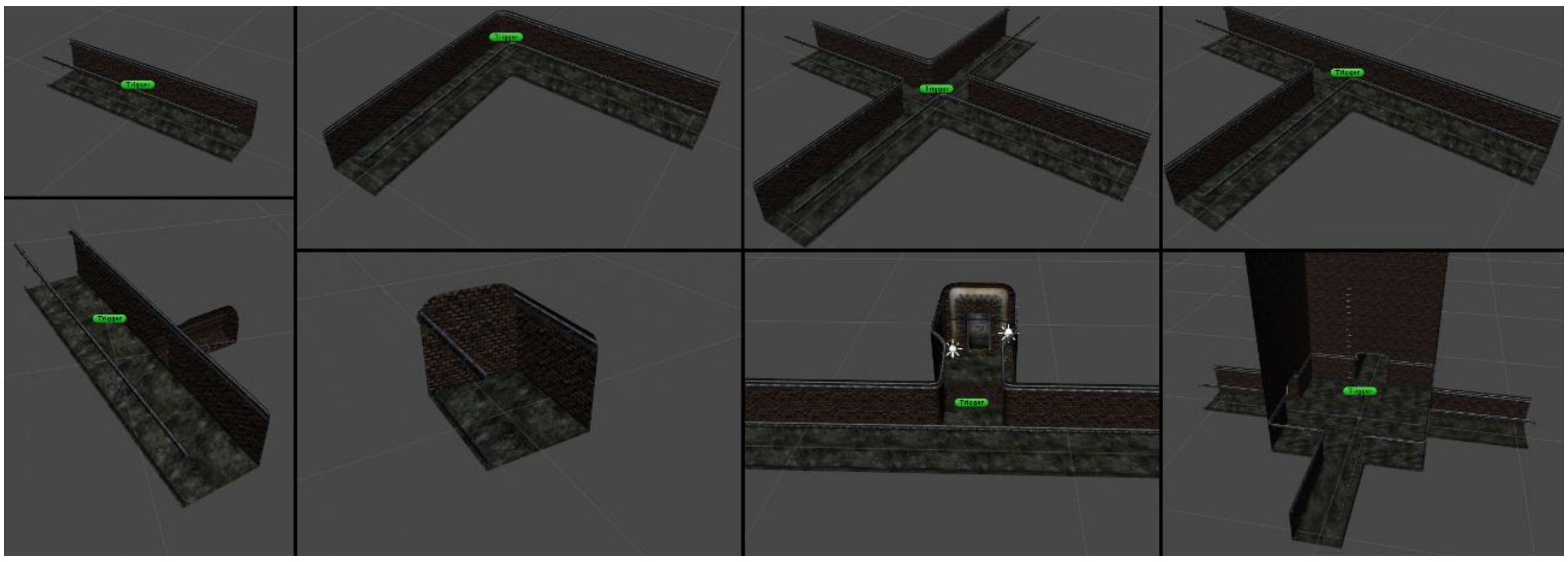

Figure 13: Different type of map blocks used for level generation.

Some of these blocks have unique properties that make their spawning mechanics different from the rest. One of these properties is a time delay that occurs after a specific block (*Y*) is spawned and prevents it from being re-spawned for *d* blocks. This means that until the block generation algorithm spawns *d* other blocks, block *Y* cannot be spawned again. On Table 3, we can see the different delays assigned to each block.

Table 3: Block delay parameter for the special block types.

| Block Name | Delay |
|---|---|
| Key Room 1 | 6 blocks |
| Key Room 2 | 6 blocks |
| Exit Room | 10 blocks |
| Dead End | First 4 initial blocks |

In order to maintain the flow and purpose of the game, we had to associate generation rule sets with certain block types. This means that some blocks cannot spawn when other blocks are already on the game world. These rules are displayed on the Table 4.

Table 4: Rule set for each special block type.

| **Block Name** | **Rule** |
|---|---|
| Dead End | Cannot spawn if there is already a Dead End block spawned. |
| Key Room 1 | Cannot spawn if at least one of the following blocks is currently spawned: Key Room 1, Key Room 2 and Exit Room |
| Key Room 2 | Cannot spawn if at least one of the following blocks is currently spawned: Key Room 1, Key Room 2 and Exit Room |
| Exit Room | Cannot spawn if there is already an Exit Room block spawned. |

As we have seen above, there are blocks with distinguished shapes. In order to make them connect with each other we had to specify the following constraints: *1)* every end of a block has to be in a standard format (size, matching materials, objects, lighting and remaining assets), *2)* they should be able to be spawned with different rotations, and *3)* the block total size would have to obey a certain scale – so if, for example, 4 corner pieces are spawned consecutively, they would have to match perfectly, otherwise the blocks would overlap each other.

When a block spawns, it will have to know where to attach to the previous block. For that, on each end of a block, there is an “anchor” object that indicates where other blocks should connect to that block and what orientation they should have (Figure 14).

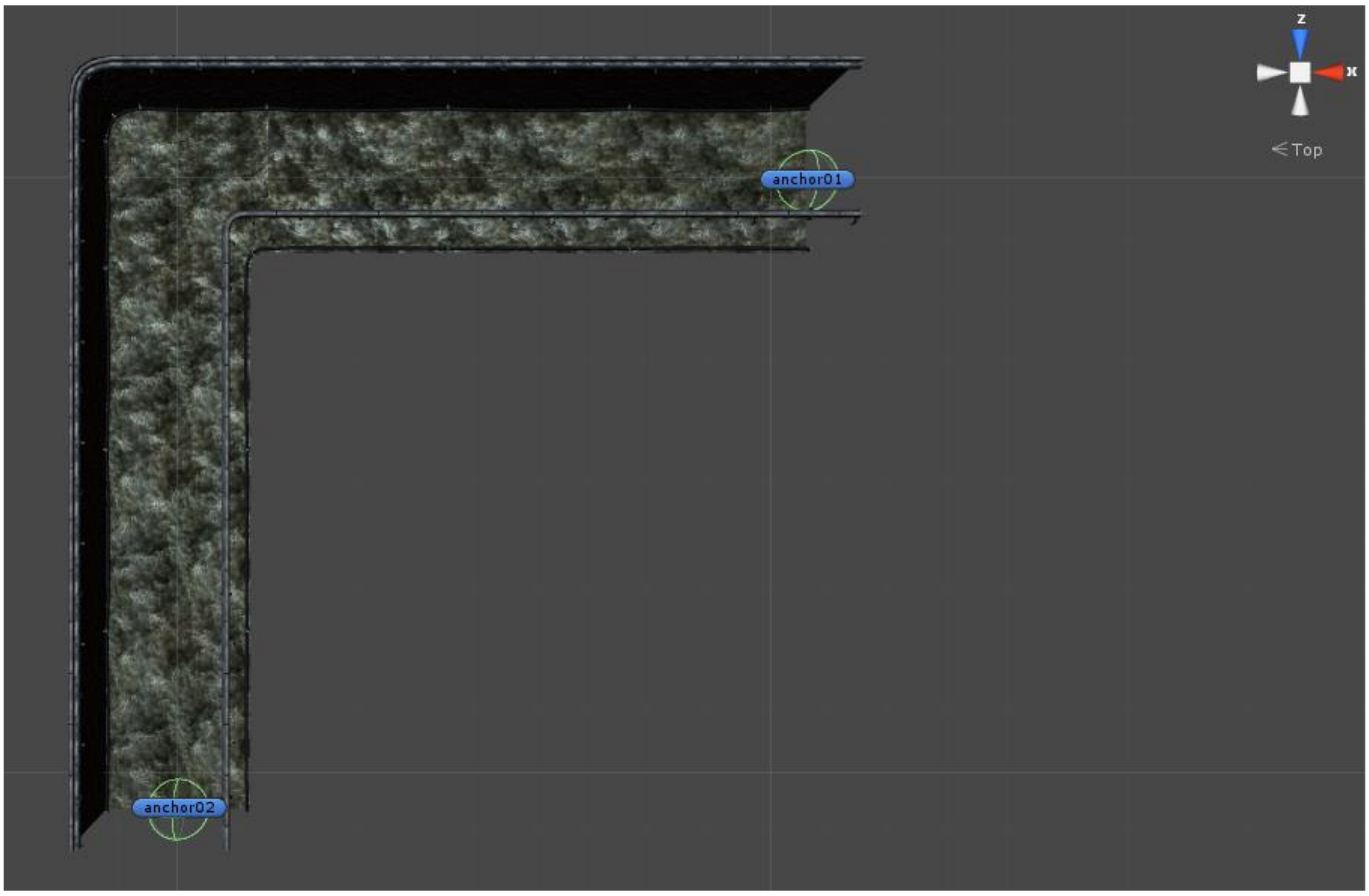

Figure 14: This image captures the anchor objects that are used for the connection between blocks.

When the game starts, only the player and the Exit Room are spawned, with the game unfolding from thereon. Throughout the game, the player has an invisible sphere around him/her, which the game uses to keep track of a list of currently visible map blocks. At each tick of the procedural generation algorithm, the game uses this list to know when, how and where blocks should be spawned and which can be safely destroyed. Whenever the player-spawning sphere reaches the centre of a block, it will set that block as "active" and spawn new blocks on the "anchor" objects that do not have any block linked to (Figure 15).

The destruction of blocks happen in a similar fashion; when the player-spawning sphere leaves the centre of a block it sets the block as "inactive" and destroys the adjacent blocks, except the one that is inside the player spawning sphere.

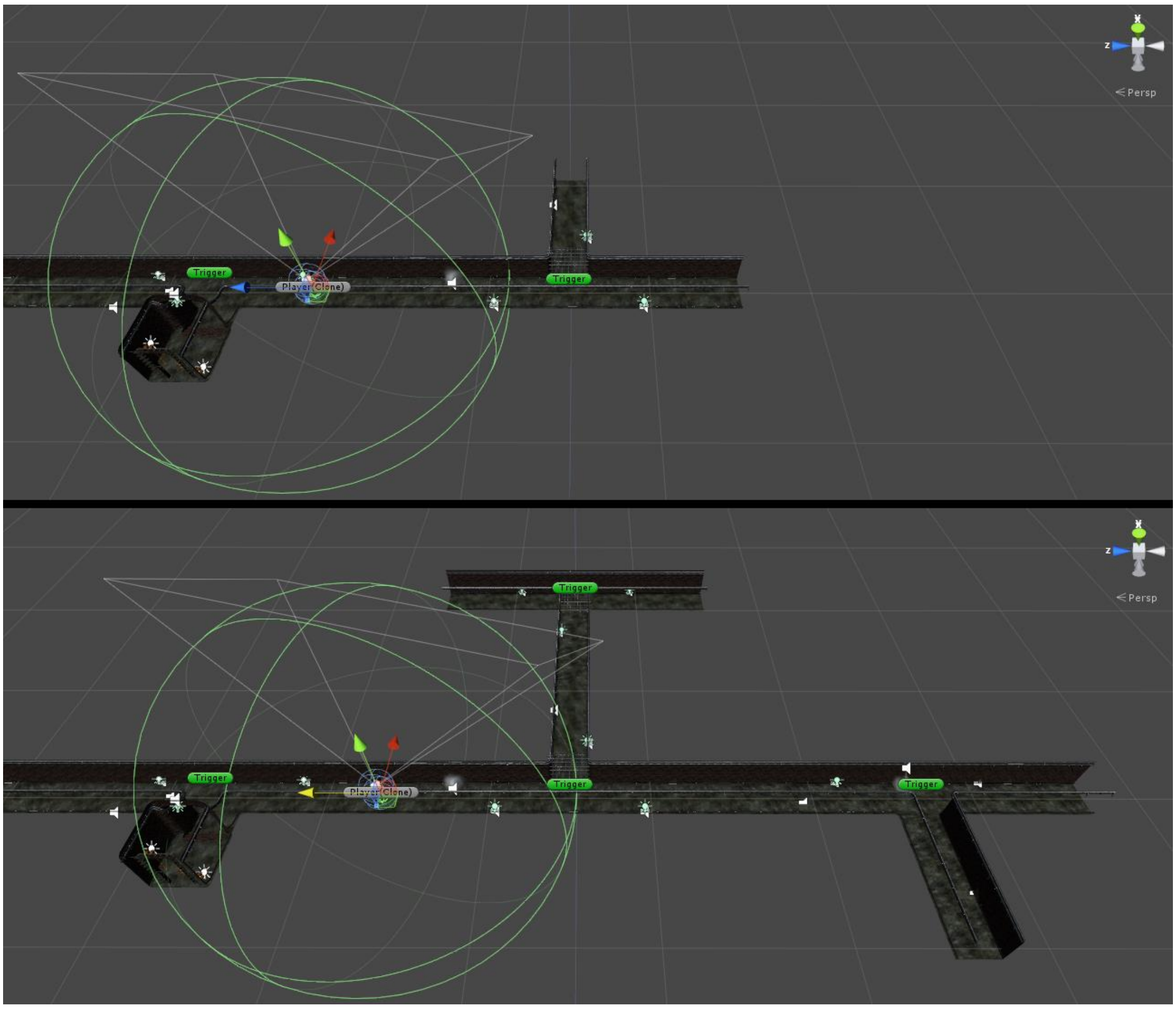

Figure 15: Scene view with the spawning sphere mechanic visualization. Top: moment before the sphere enters a new block. Bottom: new blocks spawned after the sphere triggered the spawning mechanism.

With all these mechanics we bring a new level of experience to the game, and with the rules and blocks specific attributes, we make sure that errors such as two Dead End blocks spawn at the same time and trapping the player between them, never happen. Furthermore, the way this procedural generation mechanic is set, it will be completely imperceptible to the player that the game map is changing on the fly. The player will only notice any changes when he/she decides to/has to turn back – assuming he/she has the memory to recall the map's exact layout. This mechanic also creates an interesting and powerful synergy between itself and the sanity level as the player may notice that the map changes more frequently when the character's sanity is more depleted, effectively adding an additional gameplay feature.

### 3.1.4 Procedural Content Generation

Almost all mechanics inside the game work with a certain level of randomness. As we have discussed on the previous section, the game level is being continuously generated throughout the game, and we have to make sure that the player cannot perceive this generation process throughout the game. If the blocks that are spawned were always the same, the player will probably find the game boring and repetitive. In order to counter that effect, we added over 20 detail objects (Figure 16, Figure 17, Figure 18) that can spawn inside each of the previously mentioned blocks, making them almost unique for every spawn.

Since this game belongs to the psychological horror genre, it needs events to that have the potential of scaring the player. Furthermore, since the map is not constant, these events had to be made dynamic as well. The same applies to the creature's AI; its internal mechanisms have always to take in account that the map is changing and the creature has to adapt its "thought" process and decision making to it.

#### 3.1.4.1 Block Details

On each block there are a set of "detail anchor" objects. And when the block is spawned, each detail object set for generation will be spawned under the chosen detail anchor's position. These details can be separated in the following categories:

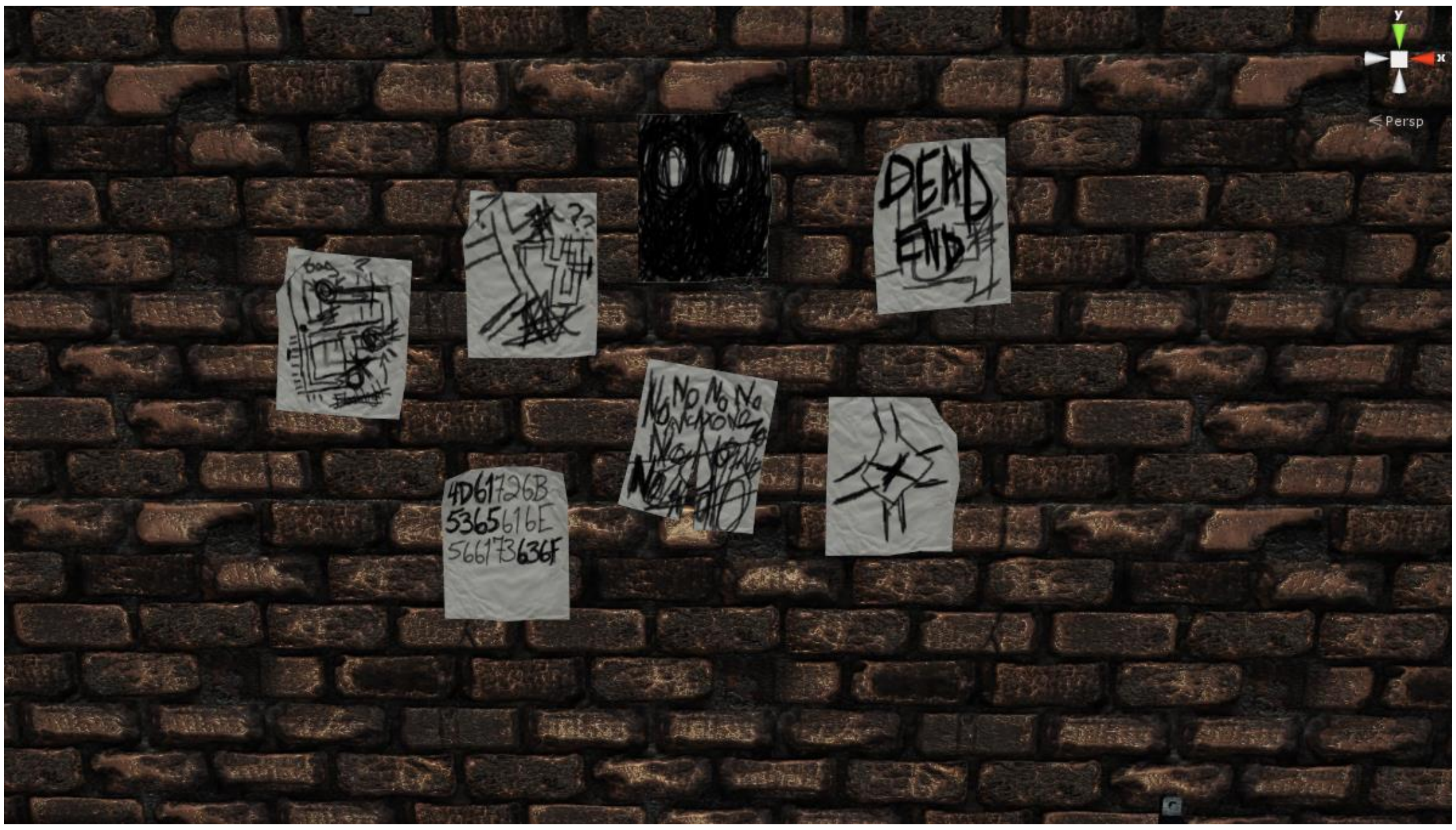

Figure 16: All the 7 distinct types of paper detail objects.

I. **Papers** - The sole purpose of the paper detail objects is to give some kind of gameplay clue. However, some of them have a twisted purpose, and they might possibly confuse the player instead of helping (Figure 16).

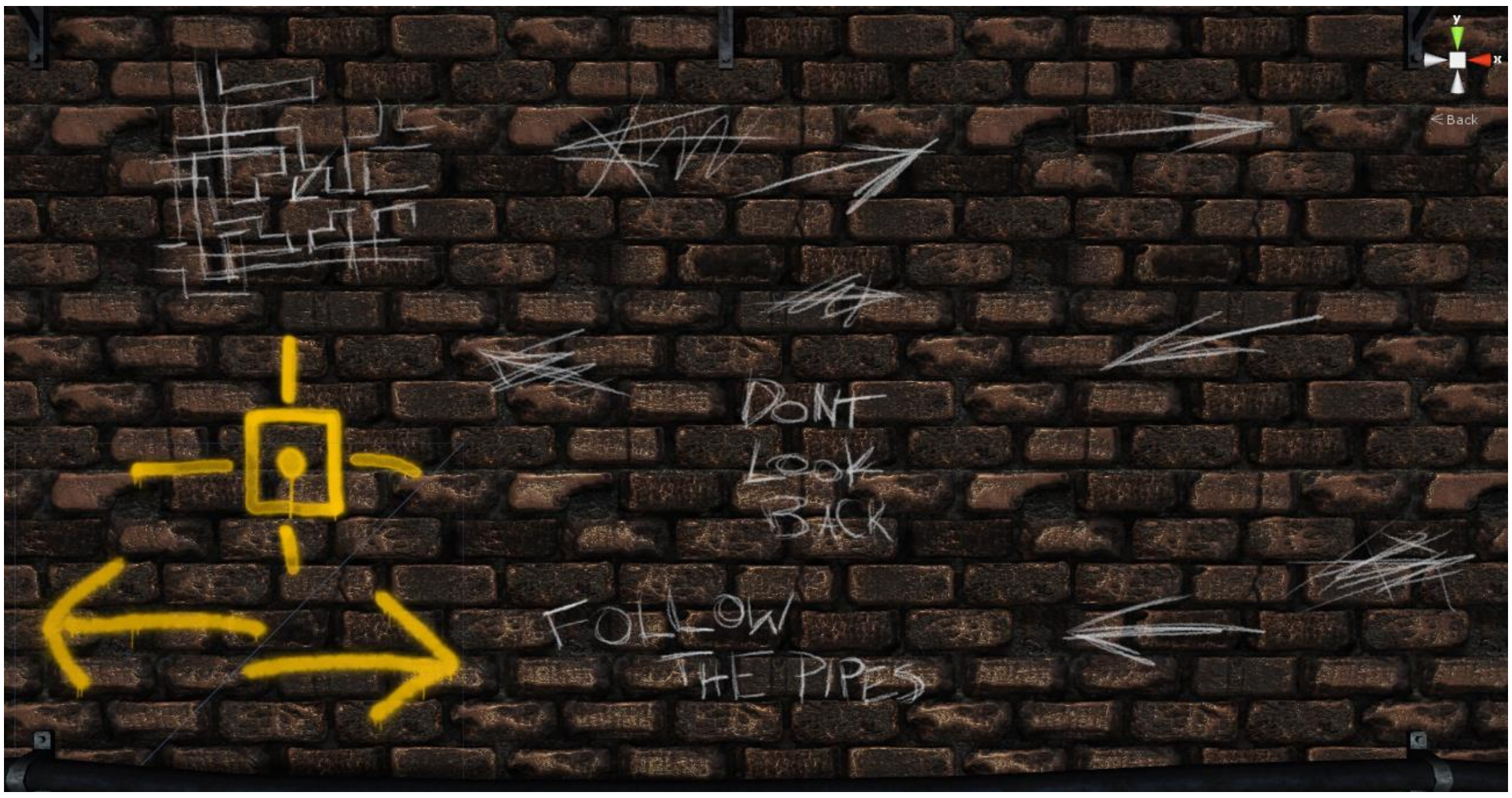

Figure 17: Showcase of all the chalk marking objects.

II. **Chalk Markings -** These detail objects' intent is to give the player some sort of aid on where he should go next. Most of these contain arrows pointing to a certain direction. But since the spawn of these detail objects is completely random there is no true aid in the game progression aspect itself, truly contributing solely to the imaginative immersion process. However, the player will only understand it later in the game (Figure 17).

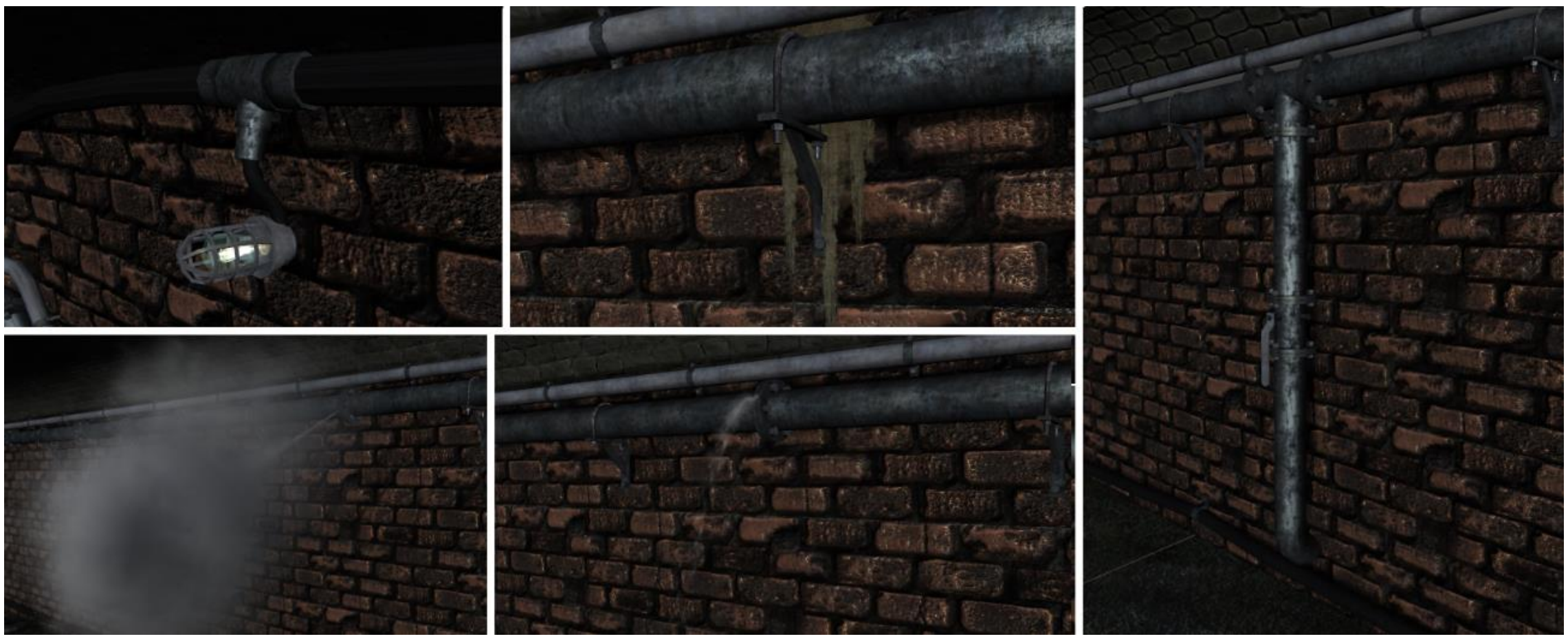

Figure 18: Different types of environmental details. Top left: wall light. Top center: water dripping. On the right: vertical pipe. Bottom left: steam. Bottom center: water splash.

III. **Environmental Details -** This set of details is a mix of different kinds of details that did not have the diversity needed to have their own set. Some of these details can be seen on the Figure 18.

Every detail object inside one of the sets described above, has a spawn time delay associated. Which means that after a specific object is spawned, he will only be able to spawn again after the delay is over. The algorithm that selects which detail will be spawned on the desired anchor checks which details are not in a delay timer and randomly picks one from that list.

#### 3.1.4.2 Environmental Events

In order to make a game more immersive, we need sudden events to catch the player unprepared. Also, because the game is being built during runtime, we had to make them dynamic. All of these events will trigger with a certain possibility when the player steps on them, similarly to a tripwire trap (Figure 19). These triggers will be attached to objects, which will be spawned inside the blocks in a similar fashion as the Block Details. These events have two main properties associated: a delay, which sets how long the event stays inactive; and a probability, that is used every time the player sets the trigger, and if the next random value from a random generator falls into the event's set probability, the event will occur and activate its delay timer. Events and their respective delays are listed on Table 5.

Table 5: List of events and their delay attributes.

| **Event(s) Name** | **Delay** |
|---|---|
| Explosion | No delay, only happens once. |
| Bugs | 20 blocks |
| Light Blast | 15 blocks |
| Pipe Steam Burst | 15 blocks |
| Pipe Water Burst | 15 blocks |
| Pipe Fall | 10 blocks |

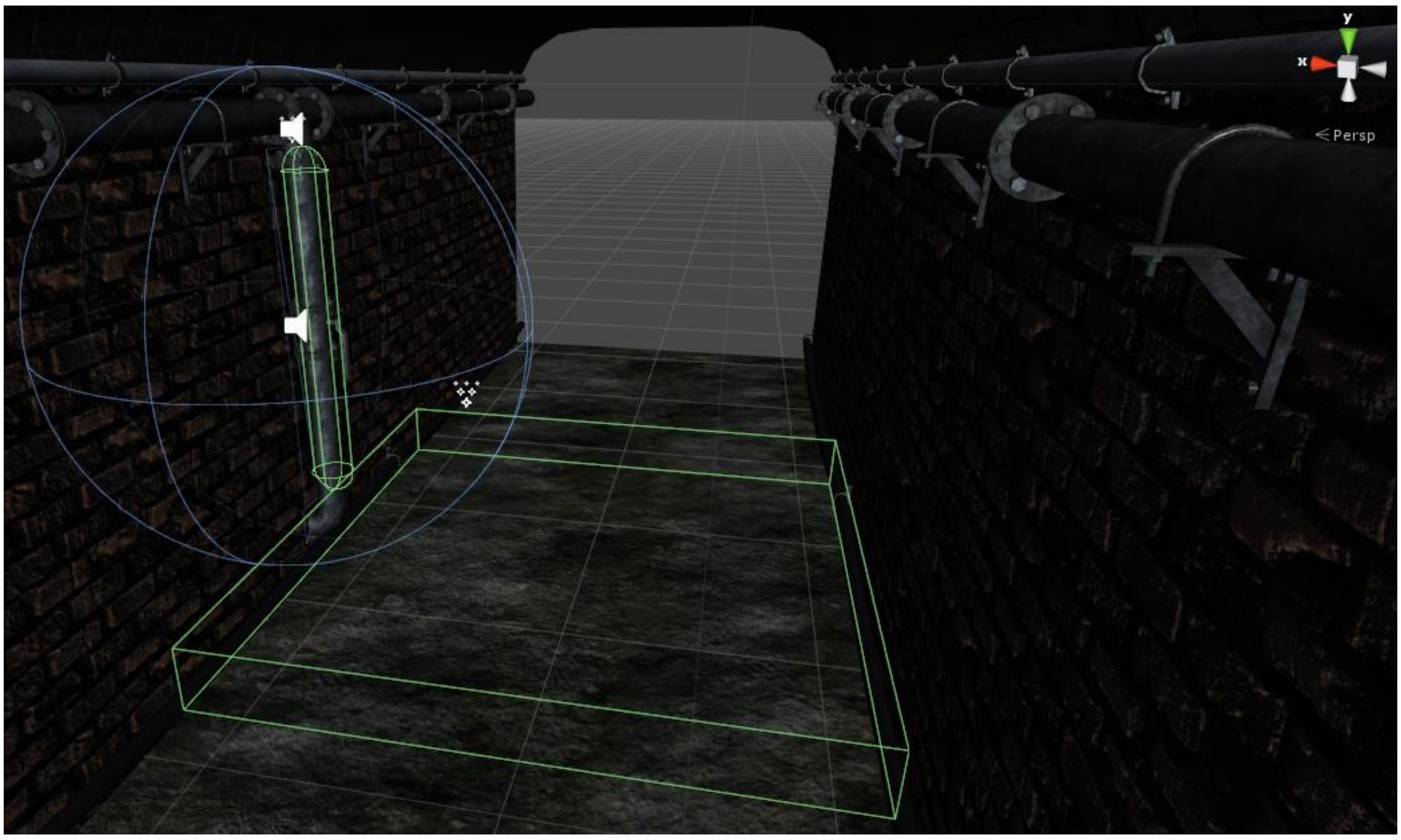

Figure 19: Game development view. The green box near the floor is the collider responsible for triggering the event.

#### 3.1.4.3 Artificial Intelligence

The game creature has the following modes of hostility:

I. **Passive** - The creature will just peek around the corners. This behaviour shows the creature's intent to stalk the player. For the player, it will work as a warning, since the creature will retreat when the player gets too close.

II. **PassiveAggressive** - This hostility mode acts as a transition phase between the Passive and Aggressive ones in order to avoid a stiff transition. The creature behaves similarly to the Passive mode, but when the player gets too close, it will now have two decisions to choose from, or it retreats like the Passive mode, or it starts chasing the player. This type might catch the player by surprise, since its behaviour will appear to be the same one as in Passive mode, but may without prior warning become incongruent with it.

III. **Aggressive** - The creature will always be hostile towards the player. It will be hiding behind the corners, hidden from the player and waiting to ambush him/her.

For the player this is the most frightening type, since they cannot predict where it will be.

When the game starts the creature's AI mode will be set to Passive, and it will change to other modes on the following situations:

- When the player picks up the first folder, the creature will enter its PassiveAggressive mode.
- Once *a)* 30 map blocks spawn and *b)* if the creature is not already in Aggressive mode it will change into it.
- After the player picks the second folder, the creature will automatically change into Aggressive mode.
- When the creature is in Passive or PassiveAggressie mode and retreats, it will increment a variable "creatureRetreats". When this variable reaches 3, the creature will go into Aggressive mode.

The creature AI state machine is composed of 4 states:

I. **Passive** - In this state the creature is only playing an idle animation while peeking around a corner. Once the player gets close enough it will then check what the creature's AI mode is. If it is in PassiveAggressive mode, it will go to the Chasing state; if it is in Passive mode, it will play a retreat animation and de-spawn.
II. **Searching** - This state is the idle state where the creature is in Aggressive mode. The creature keeps looping an idle animation and when the player gets close it will change to Chasing state.
III. **Chasing** - Here the creature will go after the player and it can have several outcomes:
    a. If the player gets far away from the creature, the creature will de-spawn;
    b. If the player hides himself in one of the evasion tunnels, the creature will go to Retreat state;
    c. If the creature gets within attack range, it will then trigger the death animation, and the player loses the game;
    d. If nothing of the above happens, the creature will keep walking towards the player.
IV. **Retreat** - In this state the creature will ignore the player and move to the centre of the block ahead of it, when it gets there it will then go into Searching state, waiting to ambush the player.

Another property of the creature is its spawning probability, which starts at a low percentage, and increases towards the end of the game, depending on the amount of blocks that have spawned.

The game map is changing at all times, which means that the creature spawn cannot be done on a global position; it has to be attached to a certain block. Every block has a list of "creature anchor" objects, which basically define creature-spawning positions. Whenever a block is spawned, if the creature is not currently positioned on the game, the block will run an algorithm and choose the best "creature anchor" to position the creature. If the next random value from the random generator falls into the creature spawning probability, the creature will spawn on that location.

In case the creature spawns in a specific block, but the player does not enter that block and instead moves farther away, when the block de-spawns, the creature will de-spawn with it.

### 3.1.5 Other Game Mechanics

#### 3.1.5.1 Sanity System

The game character is imbued with a sanity profile. When events start to happen, the character will become more and more agitated and slowly approach insanity over time. For the player to perceive this character behaviour some game mechanics were changed and a few new ones added. This sanity system as the following levels:

- **Level 1 (Sane) -** This is the starting level, no visible alterations to the character's psyche occur;
- **Level 2 (Scared) -** When the character reaches this level, his breathing will become faster and heavier;
- **Level 3 (Terrified) -** The character starts hallucinating and hearing strange sounds behind him;
- **Level 4 (Insane) -** When the character gets to this level, he will become dizzy and the player will have a hard time controlling the camera movement. On this level, the character's hallucinations will become more severe, and sometimes he will start seeing bugs on the floor and walls.

The transition between the sanity levels will happen anytime an event happens, either if it is an environment event (light bulb burst, pipe fall, etc.), or when a creature appears.

#### 3.1.5.2 Fear-Level System

Whenever an event happens, some effects will play accordingly to the intensity of the event. These effects consist on the combination between a vignette effect and the increase of the character's Field of View (FoV). This combination's intent is to simulate a tunnel vision, where its intensity will depend on the severity of the event. If it is an environment event, the effect will have low intensity; on the other hand if it is a creature event, the intensity will be higher. The fear system has also another effect, which is played only when the creature is chasing the player and the player looks at the creature. This effect involves a modification of the tunnel vision effect, and the addition of a camera shake effect.

#### 3.1.5.3 Player Controls and Mechanics

The game controls are similar to a traditional First Person Shooter game. The player can use the “WASD” keys to walk, the Shift key to sprint, and the Control key to crouch. To pick up folders the player needs to look at them and press the Left Mouse Button (LMB).

As far as the player mechanics go, the player moving speed changes according to its movement mode. If the player is running the speed increases (to x% of his base speed), if the player is walking while crouched, the speed decreases (to y% of his base speed).

## 3.2 $E^2$ Architecture

The Emotion Engine ($E^2$) is a conceptual game design framework developed with the main intent of creating a biofeedback game development standard. The architecture is directly connected to Pedro Nogueira’s work on emotional regulation and was created during his PhD thesis. Given the close working relationship between our two parties and the direct connection to the work described on this thesis we chose to adopt this framework as the means through which to enable biofeedback mechanisms in VANISH. However, since the $E^2$ architecture describes how biofeedback systems can be implemented in a general fashion and targets more complex DAB mechanisms, a direct implementation is not advisable given the limited time resources present in a Master’s thesis. As it stands, there was a need for each and every component described in the architecture. However, not all of them needed to work in real-time ($AR^ES$), so they were developed in an offline standalone fashion. The framework is divided into 4 sub-systems (PIERS, $ARE^2S$, CLEARS, GLaDOS), each one responsible for a specific task but with a high-enough degree of complexity that required its own dedicated system (Figure 20).

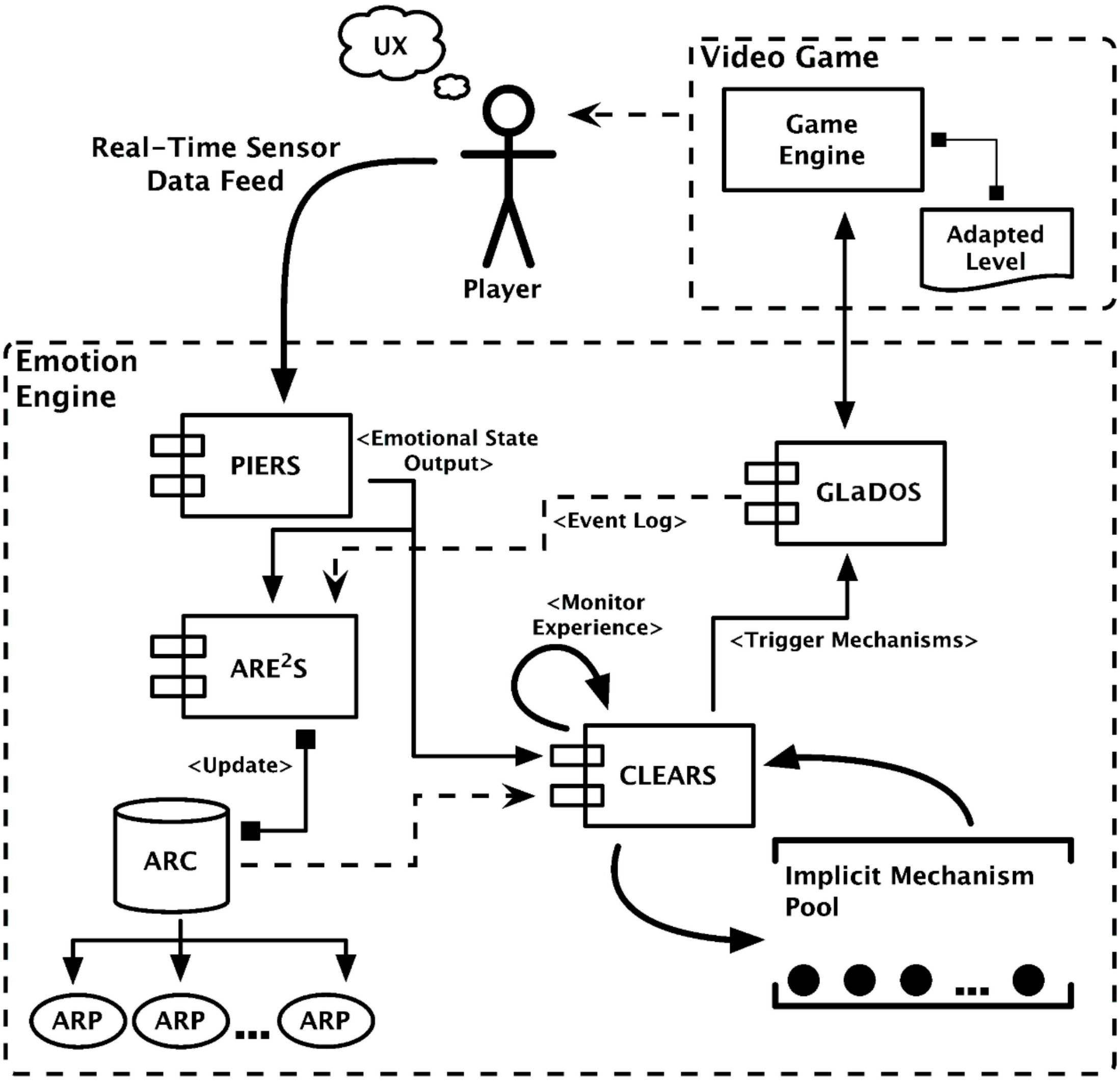

Figure 20: High-level overview of the $E^2$ architecture.

### 3.2.1 PIERS

The main data type of this architecture is the player's emotional state (ES). For this system, we adopted one of the most popular dimensional (i.e. quantifiable) interpretations of emotions – Russell's arousal and valence dimensional theory.

For our emotion recognition module we used a C# re-implementation of the Physiologically Inductive Emotion Recognition Sub-system (PIERS), as initially described in (Nogueira et al.), later implementing the real-time improvements suggested in (Nogueira, Rodrigues, and Oliveira.). Details on how this system was developed are outside the scope of this thesis and, as such, we refer the reader to the aforementioned references. In the interest of completeness however, a brief, high-level description of how PIERS operates follows.

The re-implemented version of PIERS relies on a two-layer classification process (Figure 20) to classify Arousal and Valence based on four distinct physiological sensor inputs: Skin Conductance, Heart Rate and facial Electromyography measured at the Corrugator Supercilii (brow) and Zygomaticus Major (cheek) muscles. The first classification layer uses several regression models to normalize each of the sensor inputs across participants and experimental conditions, while also correlating each input to either Arousal or Valence. The second classification layer then employs a residual sum of squares-based weighting scheme to merge the various regression outputs into one optimal Arousal/Valence classification in real-time, while maintaining a smooth prediction output (Nogueira, Rodrigues, and Oliveira.).

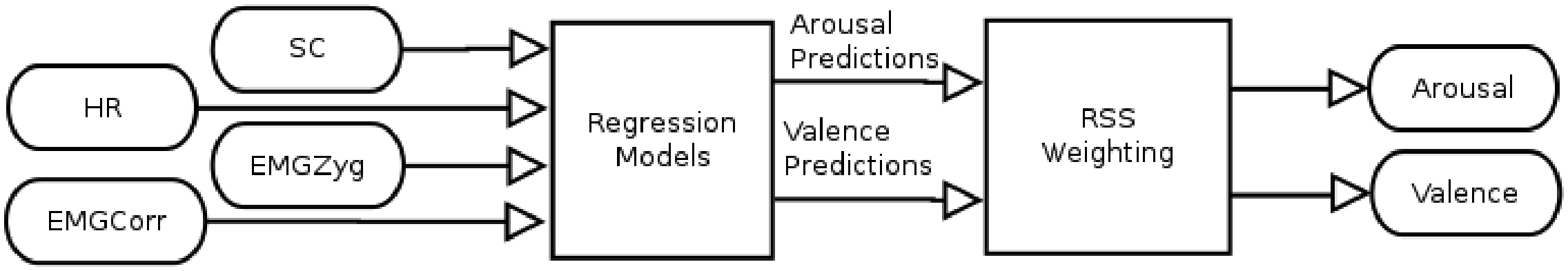

Figure 21: High-level overview of the method's architecture. Each of the physiological metrics is used to create two distinct sets of arousal and valence predictions. These predictions are then fed in parallel to a residual sum of squares-based voting system, which combines the prediction sets into one final prediction for either Arousal or Valence, depending on the provided input.

Using this methodology, the authors have achieved convincing accuracy ratings – 85% for Arousal and 78% for Valence, using 10-fold cross-validation –, which represent only a marginal (~5%) decrease in the system's previous offline implementation (Nogueira et al.). Furthermore, PIERS performs the whole emotional state classification process in a subject-independent fashion, all the while successfully predicting these states in a continuous, real-time fashion and without requiring arduous or time-consuming calibration or parameter tuning procedures. All of these features, combined with again our personal close working relationship meant that PIERS was not only suitable for our particular needs but also simplified and accelerated the implementation process, which was a critical factor in our final choice.

### 3.2.2 ARE$^2$S

ARE$^2$S stands for Affective Reaction Extraction and Extension Sub-system and it is accountable for extracting the players' phasic (i.e. immediate) emotional reactions to occurring events. Even though this component is part of the E$^2$ architecture, as discussed on this sections' introductory paragraph, it was not necessary for the required purposes and, as such, was not implemented in real-time. However, since we wanted to examine participant's responses to gameplay events to test their efficacy and also debug the PIERS system, the core components of the ARE$^2$S sub-system were implemented in an offline standalone software tool that later evolved into a psychophysiological data annotation tool for videogame user research (Nogueira, Torres, and Rodrigues). A detailed clarification with purposes and features of this tool can be seen on Emotion-Event Triangulation.

### 3.2.3 CLEARS

This component is responsible for making the decisions regarding which mechanics should be triggered based on the current player's AV rating. Once the data from the player current emotional status is transmitted from PIERS to CLEARS, the latter will then decide what action the game is going to take. This action will depend on a set of pre-established rule set. These rules have the purpose of using a game mechanic to induce a reaction on the player emotional status. During this thesis' initial brainstorming sessions we defined various possible gameplay adaptations scenarios. However, not all of them were viable on the limited time budget, so we decided to maintain only two of them. Since these gameplay adaptation mechanisms are fundamentally different, we decided to implement two independent versions of CLEARS, each one with its own rule set reflecting our visions of said mechanisms. Over the next sections we will describe these mechanisms and the objectives which precede them.

#### 3.2.3.1 Non-Visible Indirect Biofeedback

The Non-Visible Indirect Biofeedback (NV-IBF) implementation of CLEARS has the main purpose of changing the game mechanisms on a way that is not perceptible by the player.

This is the most common and obvious type of biofeedback gameplay adaptation. However, since by nature it needs control over spatial generation mechanisms (such as the procedural ones employed in VANISH) and dynamic event triggering, practical implementations are extremely rare, if not at all virtually non-existent. In our approach, the involved mechanisms control not only game features such our Procedural Generation Algorithms, but also the game's AI. The subliminal goal of this implementation is to induce emotional reactions that lead the player towards a balanced (neutral) emotional state. One parallel aspect that we were keen on finding was whether keeping the player call (and transitorily cool-headed), would help their in-game performance. The answer to this particular question can be found on Results and Discussion. On the sections below we describe the modifications that were done to the game mechanisms.

### Environmental and Creature Events

As we learned from previous alpha testing, creature events significantly increase the players' Arousal levels. Given this fact, when the players' Arousal level is high, we lower the probability of the creature appearing. If, on the contrary, his Arousal level is low, we increase the probability.

Since we still need events to happen in order to maintain the player immersed. We decided to have the environmental events behave opposite to the creature events. So, if the player Arousal is high, we increase the probability of environment events (since the probability of a creature event is lower), and if it is low, we lower the probability of environmental events. We also acknowledged that this might create heightened Tension levels and, with it, lower Valence ratings so care was taken to factor this into the Valence response mechanisms described below.

### Procedural Level Generation

On this implementation of CLEARS, this is the only mechanism that uses the player Valence level. When the players' Valence is low, the level generation algorithm will increase the probability of the objective rooms appearing. On an opposite fashion, if the Valence level is high this probability decreases. If the player is in the early stages of the game, and he/she has not picked up both folders yet, the algorithm will change the probability of appearing a Key Room. In the case where the player has already picked both folders, then the algorithm will change the probability of the Exit Room according to the player Valence level.

There is also a variation of probabilities when the creature is chasing the player. As we have mentioned before, it is possible to hide from the creature inside an evasion tunnel. Which made us modify this mechanism with the player Valence. In the case where the player Valence is low and the creature is chasing him, then the algorithm will increase the probability of appearing an evasion tunnel. If the player Valence is high, meaning the player is probably relaxed and confident on his abilities, then we decrease the probability of spawning an evasion tunnel.

#### 3.2.3.2 Visible Indirect Biofeedback

This second implementation focuses on changing game features that are more noticeable by the player. The purpose of the Visible Indirect Biofeedback (V-IBF) is to "symbiotically" connect the player to the game character, so that the player can see his/hers emotional state reflected in the game's character, possibly aiding in strengthening an empathic connection between the two. In order to achieve this purpose, we decided to change the game mechanics that are in some way related to the character. Below are the modifications that made in this implementation.

**Character Mechanics**

We decided to change the character movement speed and stamina according to the player's Arousal level. If the Arousal level is low, the character's speed will decrease, but he will be able to run for a longer period of time (i.e. a calm, rational jog that maximizes the distance travelled over time). As the player's Arousal rises, the character has more adrenaline pumping through his veins and can therefore run faster, but for an inversely shorter duration (i.e. a panicked sprint that can be used to quickly escape danger).

We also added new components to the game mechanics for this implementation. The first was the character's heartbeat, which gets stronger and louder the higher the player Arousal level is. The other feature we added was a "Faint" event, which occurs when the player's Arousal level gets to an extreme value (9.5, or higher). When this happens, the character will stumble forward and fall to the ground, while passing out. The screen then fades to black for a few (2-3) seconds and the character wakes up from a seemingly short blackout. As an added penalty, the creature may have spawned near the character, signifying it found the player while he was blacked out and then act depending on its current AI mode.

**Sanity System**

The changes on this system focused on the character's breathing and hallucinations. On the original version of the game, the sanity level controlled the character's breathing. In this modification we shifted this control to the player's Arousal level. When the player's Arousal level exceeds his neutral level, the character's breathing will pass from normal to scared, shivered breathing.

Regarding the hallucinations effects we changed when the visual hallucinations happen. Previously the bugs on the walls and floor were only visible when the character became completely insane. In this implementation, we changed it so that whenever the player's Arousal level is very high (8 or higher), or the Valence level gets really low (2, or less), this event starts to occur. When Valence or Arousal values move away from these levels, this effect will cease to appear.

**Fear-Level System**

We also decided to change when the tunnel vision effect occurred. Previously, it would happen whenever an event took place, and its intensity would be correlated with the intensity of the event. In this implementation, we made this effect controllable by the player's Valence level. When the player's Valence level drops below its neutral value, the effect takes effect. The intensity of this effect increases in an inverse proportion to Valence.

### 3.2.4 GLaDOS

The Game Layer alteration Daemon Operating Script (GLaDOS[7]) is the single framework sub-system that has to be built inside of the game (or implemented using tools that allow external applications to communicate with the game in runtime). Its purpose is to receive the orders coming from CLEARS and execute the desired actions on the game mechanics. To do so, it can – and should – be implemented as an independent layer inside the game, that is able to take over control of the game mechanics whenever necessary.

Since we have full access to VANISH's source code, we decided to integrate this layer into the game's native code, leaving all game mechanics to work normally when GLaDOS is not overriding their control protocols. However, when GLaDOS become active, it can change the game mechanics and variables, triggering its modifications in real-time. By natively implementing GLaDOS within the game's source code, we make sure that there are no delays internal or external bottlenecks and know with precision when each event happens.

The other requisite of this system is the transmission of the event logs to ARE$^2$S sub-system. So that it can then process the data and obtain a precise reaction of the player emotional state to a particular event. Since ARE$^2$S is implemented in an offline fashion, we only concerned ourselves with providing it with an accurate and as detailed as possible log of each occurring event, which, once again, due to GLaDOS's native implementation was mostly straightforward.

[7] It is also, according to Pedro Nogueira: "*A whimsical reference to a fictional psychotic and manipulative AI system that features as the main antagonist in the Portal videogame series made by Valve Software. It was also was meant as 1) a popular culture joke, 2) an acknowledgement of the influence and insight his chats with Mike Ambinder (an experimental psychologist at Valve that works closely in the biofeedback field) had and 3) a subliminal warning of the ethical concerns this type of research arises*".

# Chapter 4

# Emotion-Event Triangulation

Current affective response studies lack dedicated data analysis procedures and tools for automatically annotating and triangulating emotional reactions to game-related events. The development of such a tool would potentially allow for both a deeper and more objective analysis of the emotional impact of digital media stimuli on players. It would also contribute towards the rapid implementation and accessibility of this type of studies.

Emotion-Event Triangulation (EET) enables researchers to conduct objective *a posteriori* analyses, without disturbing the gameplay experience, while also automating the annotation and emotional response identification process. The tool was designed in a data-independent fashion and allows the identified responses to be exported for further analysis in third-party statistical software applications.

## 4.1 Requirement Analysis

Prior to starting the tool's development we conducted a series of brainstorming sessions with several other physiological researchers (N=16). Given that psychophysiological research is performed not only by computer scientists such as ourselves, but also by non-technical individuals from the social sciences and psychology fields we took special precautions to ensure all these groups were as equally represented as possible in our study. Ultimately we arrived at the following system requirements:

1) To provide a complete, yet easily interpretable measure of the volunteer's emotional state

2) To provide a real-time and synchronised view of the volunteer's gaming session from both an audio-visual and psychophysiological perspective
3) To allow free manipulation of the experiment's rate of time passage (i.e. to quickly scroll through the experience)
4) To allow for a simple and straightforward annotation of relevant events with as few clicks and parameter selection as possible
5) Present time markers for each of the annotated events and the ability to quickly cycle through and edit them
6) To allow for subjective data to be included for each event, if necessary
7) To automatically compute which events triggered emotional reactions
8) To incorporate a save/load feature for resuming the annotation process in relatively large data collections and posterior analysis/verification.

The latter requisite was added in the final stages of our focus group discussions since it was pointed out studies commonly amass several hours of data (three or more) on a single session. These sessions are difficult to reliably annotate in one pass by a single researcher. In fact, is not uncommon for multiple researchers to annotate or review each other's work in order to reduce inter-subject variance induced errors. Furthermore, several participants stressed the importance of, in addition to the audio-visual and physiological data, being able to import a list including each of the annotated events (e.g. as outputted by a game log). This list would, in theory, allow the tool to automatically annotate the whole session without any user input. It also became clear that it would be of critical importance that the software would be able to use emotional recognition methods others than our own and that their usage should be transparent to the user. Finally, the tool should be able to export the identified reactions to a structured output file, so that these could be further analysing in greater detail in the common third party statistical analysis packages (e.g. R, SAS, SPSS, Weka, etc.). As such, the following requirements were added to the initial ones:

9) To present the ability to not only import the audio-visual and physiological data, but also import a list comprising each of the annotated events (e.g. as outputted by a game engine or logging software) and automatically annotate the whole session without any user input
10) Transparent emotional classification (i.e. no *a priori* knowledge needed)
11) The ability to quickly export the identified reactions for later analysis in popular statistical analysis software (e.g. R, SAS or SPSS).

## 4.2 Tool Development

Since we aimed at developing a standalone solution that could be used freely regardless of the game engine or stimuli presented in the experimental protocol, we decided to build our tool from naught. After a brief survey of the available open source libraries and development time cost, we settled in using C#. Since our tool is meant to be applicable to a wide range of situations, it requires some parameters to be set (video and physiological data initial timestamps, emotional classification parameters, types of events and location of video, physiological and annotated event files). As such, we decided to store these values in a simple text configuration file for ease of use. Furthermore, given the modular nature of our framework, we choose to divide it into various independent components, so that any future additions or changes could be performed in an expeditious manner. These components are, in order of appearance in this paper: the physiologically inductive emotional recognition module (4.2.1), the event annotation module (4.2.2) and the emotional reaction identification module (4.2.3). Throughout this section we will cover each of the aforementioned modules, how they work and which features they comprise.

### 4.2.1 Emotion Recognition Module

For our emotion recognition module we used and adaptation of the PIERS system. Since its adaptation was not too significant, we will not give much focus here, see 3.2.1.

### 4.2.2 Event Annotation Module

Since a considerable proportion of our requisites (55%) were related with how to visualise and annotate the recorded material, we devoted a great deal of attention to the development of the event annotation module (EAM). Its function is to address the requirements related to the annotation functions (2-6 and 9). To fulfil requirements 1 and 2, we decided to combine a custom video player and time series graph drawing library (see Figure 21). Timestamps were logged for both the gameplay videos and the physiological data. These were then used to synchronise the emotional classification and gameplay video streams. The video player component was designed to allow the user to quickly skip through the video using a simple slider or to accelerate the video through a fast-forward and backwards button. The system was later tweaked to allow the user to skip through the data using the emotional classification time plot (see Figure 21), by simply clicking on the region of interest to skip to that point in time. This was done to improve the tool's usability as sometimes the emotional classification reveals interesting events that might be missed using the video.

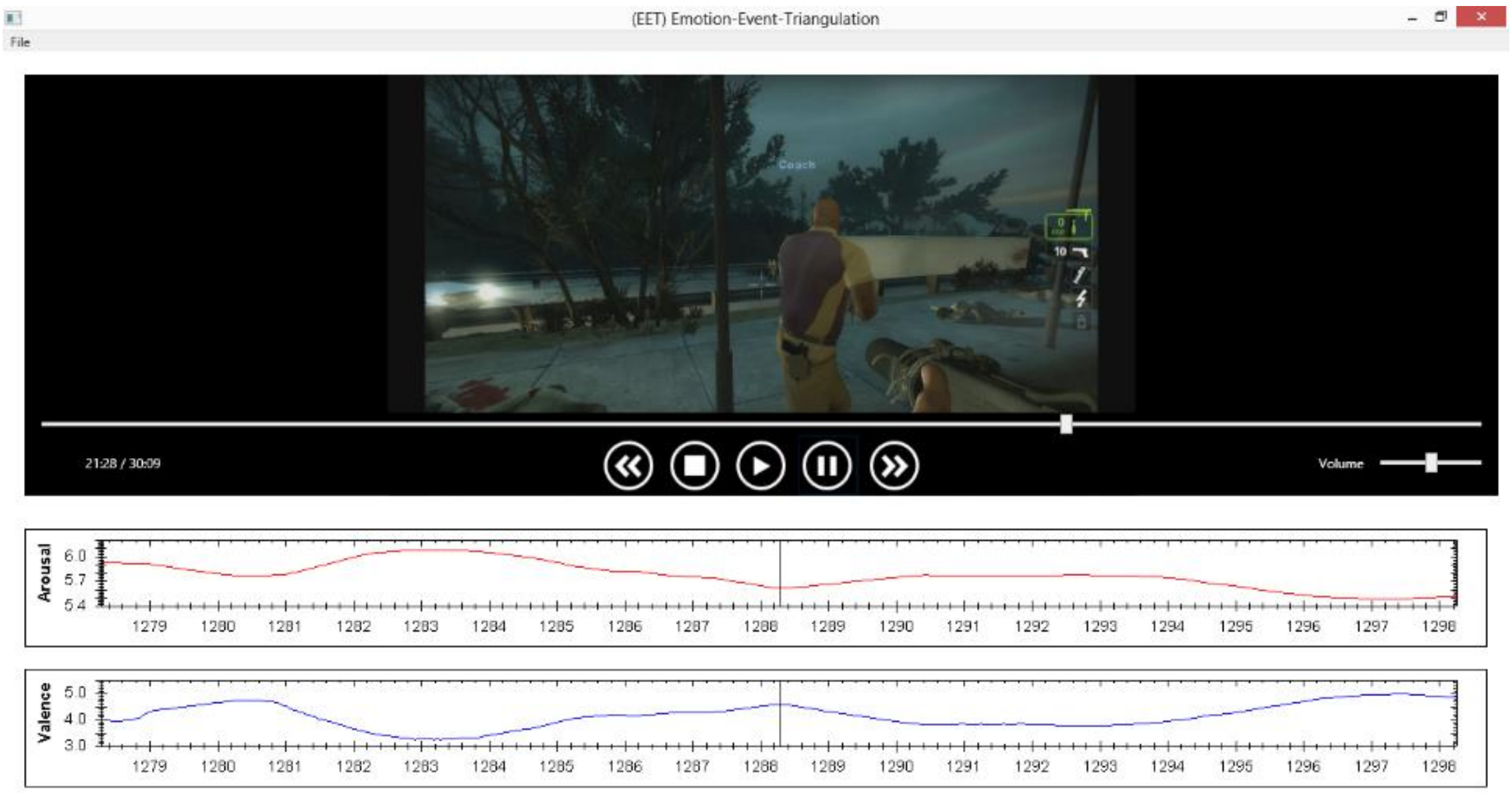

Figure 22: A screenshot of the EET tool showing the video player controls and emotional classification time plot.

Regarding the event annotation process itself (requisites 3-5), we decided to limit the user input to the barest essentials in both terms of actions and input required. To insert a new event, the user can either perform a right-click on the video player window or right-click on the emotional classification time plot and choose "Add new event" (Figure 22). In both cases, this will add a new event at the current time and bring forth a pop-up form where the user can choose which event took place (if no event file is loaded) and any subjective commentary deemed relevant (requisite 6). Finally, the user can access a list of recorded events by, again, right-clicking on the video player or emotional classification time plot and choosing "Edit Events". Double-clicking on any of these events will automatically shift the user the event's timestamp and open it's parameterisation window, as if adding the event for the first time.

Finally, attending to requisite 9, we added a feature to allow importing a list of previously annotated events. This is done using an optional field to the tool's configuration file (as described in section 4.2) so that if the location for the event list text file is given, the tool automatically parses the file and loads each event. While this was not our case, this feature was added to account for scenarios where it is possible to automatically generate an event timestamp list. Using this feature, our tool can virtually allow the user to perform the annotation process in a matter of minutes by simply writing the configuration file, loading it and commanding the tool to identify the occurring emotional reactions (see the next section for further details on this process).

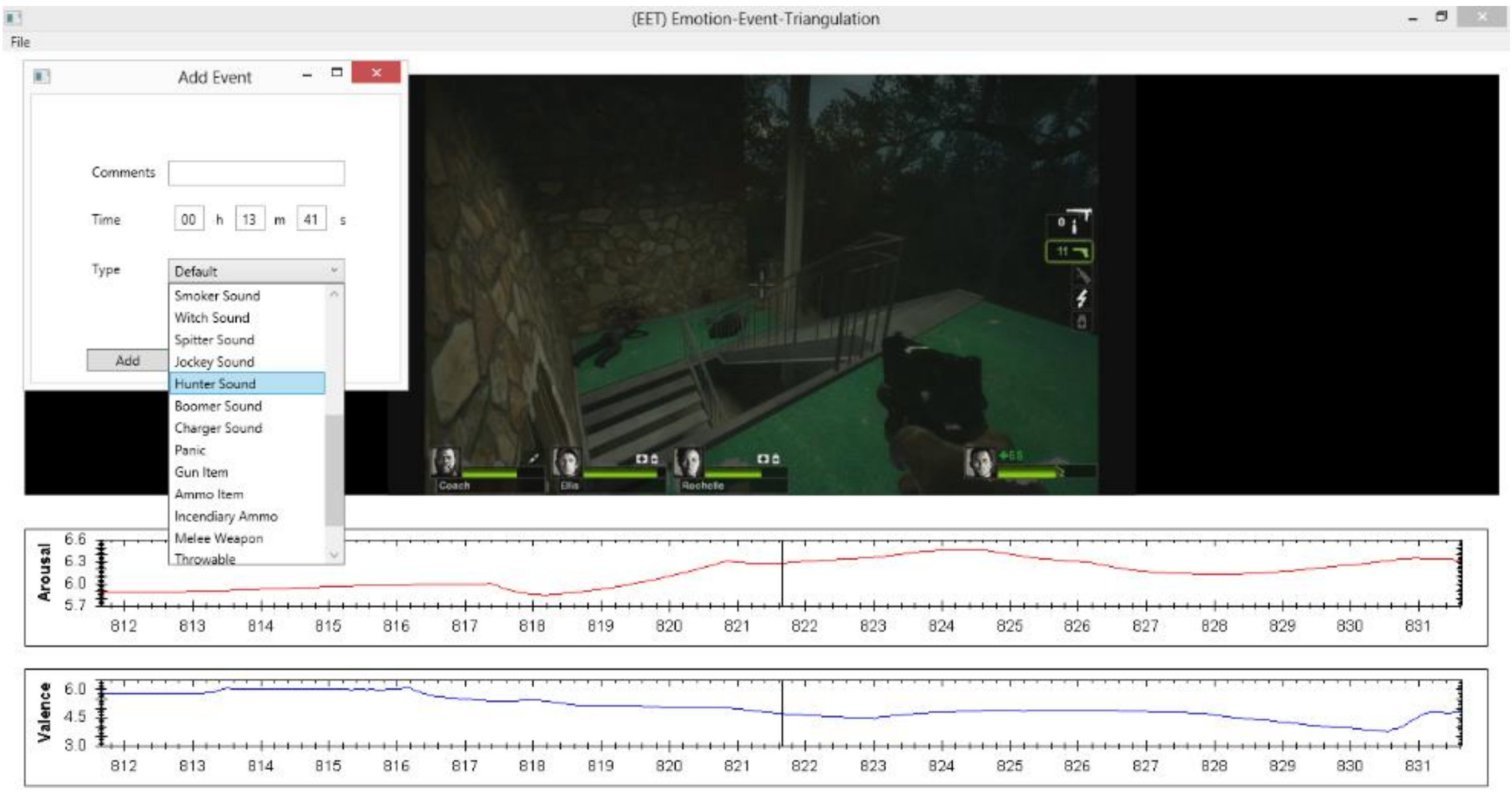

Figure 23: The add event window, super-imposed on the EET tool. Although it is editable, the time stamp for the event is automatically filled-in with the current timestamp. The user only needs to choose which event is occurring/going to occur and include any relevant comments (optional).

### 4.2.3 Emotional Response Identification Module

The final component in our tool is the emotional response identification module, which is responsible for performing the basic triangulation between the recorded events and the ensuing responses in the Arousal and Valence dimensions. Its main feature is the automatic identification of the annotated events (requisite 7). The triangulation process was automated by developing a simple local peak detection algorithm that, given an initial emotional state comprised of an Arousal and Valence ($IS_A$, $IS_V$) classification, searches forward in time to identify zero-crossings in each dimension's first-order derivative. The emotional classification values observed at each zero crossing are then compared to a minimum absolute local variability threshold φ, such that $| IS_A - ZS_A | \geq \varphi_A \vee | IS_V - ZS_V | \geq \varphi_V$ and $\varphi_A = (\mu_A + 2\sigma_A)$, $\varphi_V = (\mu_V + 2\sigma_V)$, where $ZS_A$ and $ZS_V$ represent the Arousal and Valence dimension values at each zero crossing state and $\mu_A$ and $\mu_V$ denote the mean value of the Arousal and Valence dimensions in the event's time region respectively. Likewise, $\sigma_A$ and $\sigma_V$ denote the standard deviation Arousal and Valence dimensions in the event's time region. A time region (Ω) for a particular event $e_i$ defined as $\Omega = [T(e_i), \min(T(e_i)+10, T(e_{i+1}))]$, where T is the mapping function between an event and its corresponding timestamp. In other words, a time region is the time interval spanning from the event's timestamp to either 10 seconds in the future, or the timestamp of the following event). The 10-second window was defined having in mind: *a)* the response delays of the physiological data used in our emotional classification method (up to 5 seconds for SC); *b)* the time the stimuli may take to be perceived – from empirical analysis, in average 1 to 2 seconds – and *c)* the time the emotional response may take to fully manifest itself. The 10-second window was also designed to account for multiple emotional responses to the same event – a phenomenon that was not initially expected. However, it seems some events have the capability of eliciting various (sometimes conflicting) emotional responses. For example, it is fairly common for certain enemies to elicit both low and high Valence responses, which is due to the enemy's relation to the gameplay mechanics. Such an example is the Boomer enemy, which is a large, obese character that explodes when shot or within detonation range of the player. As such, it poses both a considerable threat and tactical advantage (if detonated near a group of weaker enemies). It is understandable that when hearing the groan of this enemy type, players felt negative Valence (fearing he was close) and then positive Valence (when spotting him near a group of enemies and detonating him). To account for this type of emotional responses – which we dubbed composite responses – we further tweaked the peak detection algorithm to identify all local maxima (peaks) and minima (valleys) within the event's time region, by using the previous' response maxima/minima as the initial emotional state from which to search. The accuracy results for the peak detection algorithm can be found in the next section, along with their discussion. An illustrative example of the output provided by the algorithm can be found bellow in Figure 23.

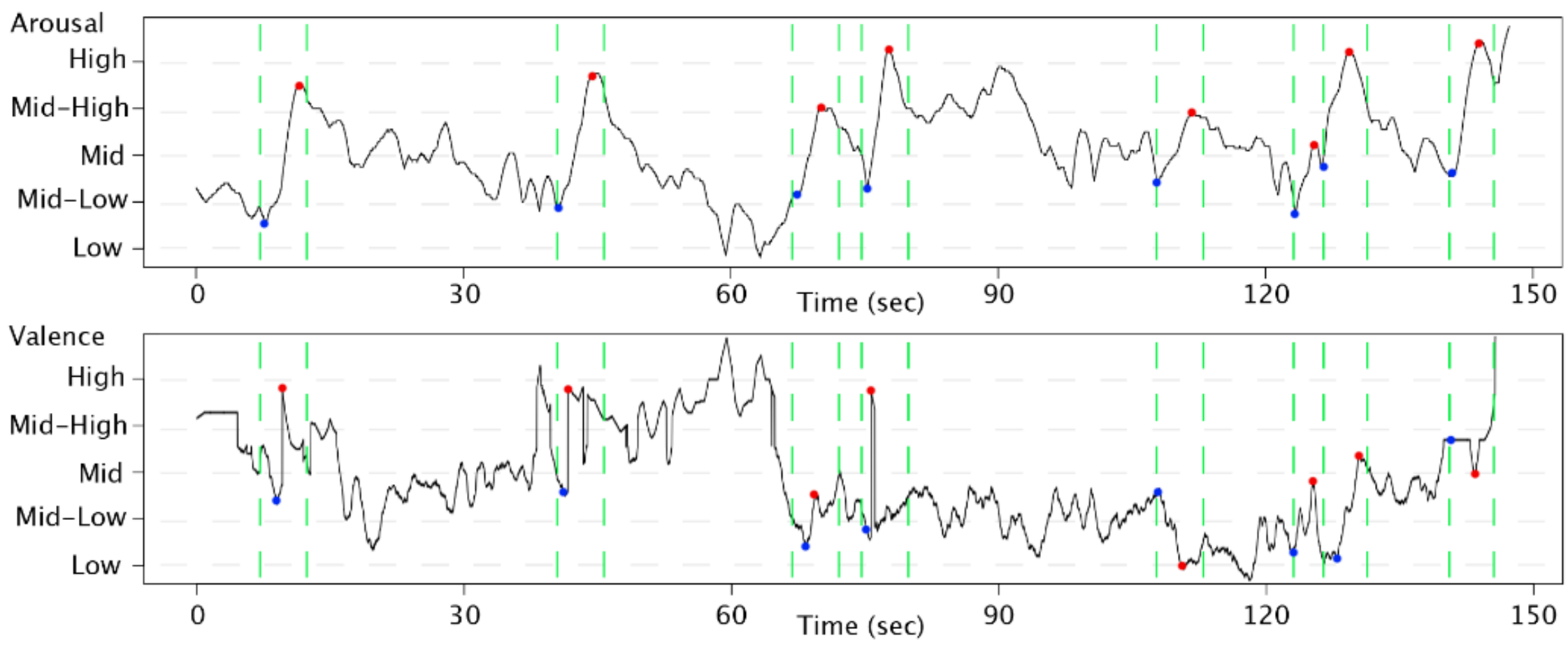

Figure 24: Example of the output provided by the peak detection algorithm over a 150-second window. The emotional output was discretised into 5 levels for both Arousal and Valence for interpretability. Assume that the distance between levels equals the local minimum variability threshold φ. Blue circles denote the timestamp for each logged event; red circles denote the identified local maxima/minima; green dotted lines represent the event's time region (set to 5 seconds for reduced complexity in this example).

To fulfil the remaining requisites (8 and 11), this component was also endowed with the ability to export the identified reactions to a structured text file for posterior analysis (requisite 11) and to serialise the entire tool's internal state to a custom *.eet* file extension (requisite 8). The latter allows the annotation process to be resumed or re-analysed in a posterior point in time. Finally, since there is no universally accepted format for physiological data storage, our tool currently accepts the format provided by the BioTrace+ software, which we used in the experiment. Since statistical analysis software solutions usually have its own input data format, we also chose to define our own unique format for exporting the identified reactions.

## 4.3 EET Results

In order to test the adequacy of the peak detection algorithm, we decided to compare the obtained results from the automatic detection to a manual approach. To this end, volunteers' gameplay sessions where randomly chosen and annotated. Since each gameplay session occurred over a relatively large time frame ($\mu = 37.4$, $\sigma = 11.4$ minutes) and a large number of events were recorded in each one ($\mu = 72.0$, $\sigma = 28.4$), this implied a considerable time effort in manually annotating each session. Thus, we decided to randomly select six volunteers and use them to validate our algorithm. Overall, a total of 430 gameplay-related events were identified and annotated, to which 364 emotional responses were observed – an average of 88.24% event/response ratio, with an 11.52% standard deviation. Out of these 364 identified emotional responses, a considerable minority of them related to simple responses ($\mu=16.9$, $\sigma=8.1$), with the remaining 83.1% corresponding to composite responses. This presented an unexpected result that, in our opinion, further justifies the latter enhancement of the peak detection algorithm to detect this type of responses, as discussed in the previous section. Pertaining the algorithm's local maxima/minima identification accuracy, it revealed to be of adequate performance, as the results present in Table 6 bellow indicate. Overall, the algorithm was able to identify local maxima and minima with a success rate of 93% for simple responses and ~94.5% for composite ones.

A response was considered correctly identified only if all maxima/minima were detected. The fact the algorithm presents lower detection accuracy for the simple response category may be justified by both the lower sample population and by its poor performance on volunteer B (whom presented very shallow peaks, which led the algorithm to ignore them and us to acknowledge them). However, it remains unclear whether these should be considered or not, as it is a subjective question warranting further investigation. Finally, it is worth mentioning that in the algorithm did not present any false positive results (i.e. detecting an emotional response where we considered none to be present). Although the algorithm's sensitivity is tunable, this was a trade-off we considered fair in terms of shallow maxima/minima detection and false positive results for this particular study.

Table 6: Number of observed emotional responses across all six randomly chosen volunteers and respective detection accuracy ratings. Simple response detection shows similar performance to composite responses, albeit with a larger standard deviation – probably as a by-product of the smaller sample and algorithm's parameterisation.

| **Volunteer Code** | **Number of Responses** | | **Detection Accuracy** | |
|---|---|---|---|---|
| | **Simple Responses** | **Composite Responses** | **Simple Responses** | **Composite Responses** |
| A | 17 | 49 | 94.12% | 91.84% |
| B | 11 | 60 | 63.64% | 98.33% |
| C | 8 | 30 | 100% | 86.67% |
| D | 7 | 49 | 100% | 94.0% |
| E | 4 | 78 | 100% | 97.44% |
| F | 12 | 39 | 100% | 97.40% |
| Total ($\mu$, $\sigma$) | 59 (9.8±4.5) | 305 (50.8±16.8) | N/A (93.0±14.6) | N/A (94.3±4.5) |

The system described in this chapter enables game UX researchers to quickly annotate game events and analyse players' emotional responses via their physiologically classified emotional states. This new analysis pipeline aimed at reducing the associated workload to the annotation process, while eliminating human subjectivity errors and contributing towards the standardisation of this type of studies.

Preliminary results suggest a habituation function may indeed be modelled from the acquired data, while the effects of the initial emotional states are yet to be verified.

In principle the system fulfilled all of the established requirements for our specific study, while retaining a generalizable approach – a feature not widely adopted in the only relating earlier work (Matias Kivikangas, Nacke, and Ravaja 2011). This versatility is dictated by the tool's independence towards the input data and by the emotional recognition module's modular design (which can easily be swapped by another implementation). Furthermore, the tool does not limit the data analysis process to its capabilities, as it allows the user to export the detected emotional responses for further statistical exploration or modelling.

However, this annotation process is not yet without its flaws, as we have identified two of them. Firstly, there is a trade-off between false positive and false negatives in tuning the peak detection's sensibility thresholds. Still, it remains to be seen if the error introduced by this trade-off is not smaller than the one introduced by human error (i.e. inter and intra-subject variability).

Despite this, the user can manually correct any automatically obtained results, which eases this issue and still results in a swifter annotation procedure. The second encounter issue is related to the tool's versatility, since it is logistically impossible to integrate it into every existing game engine, not to mention other types of applications or games without log outputs.

Current UX research methods are unable to perform in-game evaluations without disrupting – and thus potentially contaminating – the gameplay experience. Moreover, emotional state classification methods are difficult to integrate in these studies due to their complex development nature and technical skillset. The methodology presented in this chapter has the potential to contribute to a wider accessibility of emotional response studies by, not only easing the aforementioned issues, but also by removing the necessity of developing standalone emotional state detection systems – which in itself contributes to a standardisation and comparability of the annotation process.

## Chapter 5

# Tests and Result Analysis

Due to the different game conditions that were developed in order to assess the $E^2$ framework, some questions had to be answered:

I. Is there a significant difference between conditions with and without a biofeedback mechanism?

II. Does gender or game interest affect the UX toward a specific game condition?

III. The participants have different AV ratings between the various game conditions?

Chapter 5 presents the results obtained using a several types of analysis. On the sub-section 5.1 we describe the participant's information and a demographic analysis. Sub-section 5.2 presents the experimental procure. The data metrics used for the analysis are introduced on sub-section 5.3. And finally, on sub-section 5.4 we present and discuss the results.

## 5.1 Participants

Data was recorded from 24 healthy higher education students. Unfortunately due to sensor malfunction during one of the game conditions, the data from one of the participants was corrupted and could not be used. From those 24 participants, 8 (33,3%) were female, and the remaining 16 (66,6%) were male. Their age ranged between 19 and 28 (22.4782, 2.4999). The study was advertised through a dynamic email to the University's student community and the selected participants were selected randomly from the interested candidates (N=89). As part of the experimental setup, demographic data was collected. The remaining collected information not already described can be seen below on the Table 7.

All of the participants owned and played PC games, 64% played Wii or Mobile and only 43% play PS2/3 games. 67% of all participants played horror games before this experiment but only 47% actually enjoy playing them.

Table 7: Participants demographic information.

| Gaming Device Preference | |
|---|---|
| **PC** | 100% |
| **PS2/3** | 43% |
| **Wii or Mobile** | 64% |
| **Horror Game Preference** | |
| **Plays Horror Games** | 67% |
| **Enjoys Horror Games** | 47% |
| **Weekly playing time** | |
| **0-4** | 59% \| Girls (75% - 6 out of 8) |
| **4-8h** | 12% |
| **8-16h** | 18% |
| **16+** | 11% |
| **Gamer Types** | |
| **Hardcore** | 42% (10 out of 24) |
| **Softcore** | 58% (14 out of 24) |

All participants played games at least once a week, where 59% played up to 4 hours per week, 12% played from 4 to 8 hours a week, 18% played from 8 to 16 hours a week, and the remaining 11% played more than 16 hours a week. Girls were the most casual players, where 75% (6 out of 8) only played a maximum of 4 hours a week. Men were almost equally distributed.

Furthermore, we also classified the participants as hardcore (more than 4 hours a week) or soft (less than 4 hours a week) players. Through this classification process, 58% of them were categorized as soft players, and 42% as hardcore players.

## 5.2 Procedure

All experiments were conducted on weekdays between 10:00 a.m. and 7:30 p.m., with each experimental session lasting approximately 1 hour and 30 minutes. The experiments were advertised using a dynamic e-mail message across all students from Faculdade de Engenharia da Universidade do Porto.

After a brief description of the experimental procedure, each participant filled in a compulsory informed consent form, with a request not to take part in the experiment if they had any body piercings (as they could interfere with the biometric acquisition hardware) or suffered from any psychological or cardiovascular diseases.

Participants were then seated in a comfortable chair, which was adjusted according to their individual height, and then asked to take any belongings that could interfere with the sensors. Electrodes and sensors were attached. SC was measured at the subject's index and middle fingers using two Ag/AgCL surface sensors snapped to two Velcro straps. BVP was measured at the thumb using a clip-on sensor. Facial EMG was measured at the zygomaticus major (cheek) and the corrugator supercilii (brow) muscles and, as indicated in the literature, correlated with positive and negative valence, respectively (Stern, Ray, and Quigley 2001). The participants were then asked to relax and tell if they could see and move clearly with the interference of the electrodes.

After a resting period of 3-5 minutes, we initialized the calibration process, which was divided into the following phases:

- **Relaxing Music** - The participant was asked to put a pair of headphones and relax. Advice was given for the participant to close his eyes in order to obtain a better state of relaxation.
- **Waldo Scare -** The participant was asked to find Waldo (Duckett) on an image. This image was modified and had no Waldo in it. After 20 seconds the image would switch to a scary face combined with a screaming sound, with the intent of provoking an increase of the participant's Arousal level (see Figure 24).
- The following videos were chosen based on (Schaefer et al. 2010):
  - **Funny Video -** The participant was asked to watch a short clip of a known comedy film (American Pie: The Wedding) (Dylan).
  - **Horror Video -** The participant was asked a video that induced a feeling of despise (American History X) (Kaye).

During all of these phases, the participant was left alone in the room, so that his emotional state would not be influenced by our presence. After each of these phases, we would ask for the participant to relax, while we analysed the data. Subsequently we asked for the participant's own subjective rating of his/hers Arousal and Valence levels during that phase.

Once all the calibration data for the participant's regression models was fed to PIERS, the participant was allowed to start playing the first game condition. For this experiment we had each participant play 3 game conditions: Non-Visible Indirect Biofeedback, Visible Indirect Biofeedback and Non Biofeedback. The Non-Biofeedback condition was a version of the game that did not have the $E^2$ framework augmenting its game mechanics. The order that the participants played these conditions was randomized to prevent an habituation effect on the results. This is also correlated with our choice of having 24 subjects, so we could have the same amount of participants with identical order of conditions.

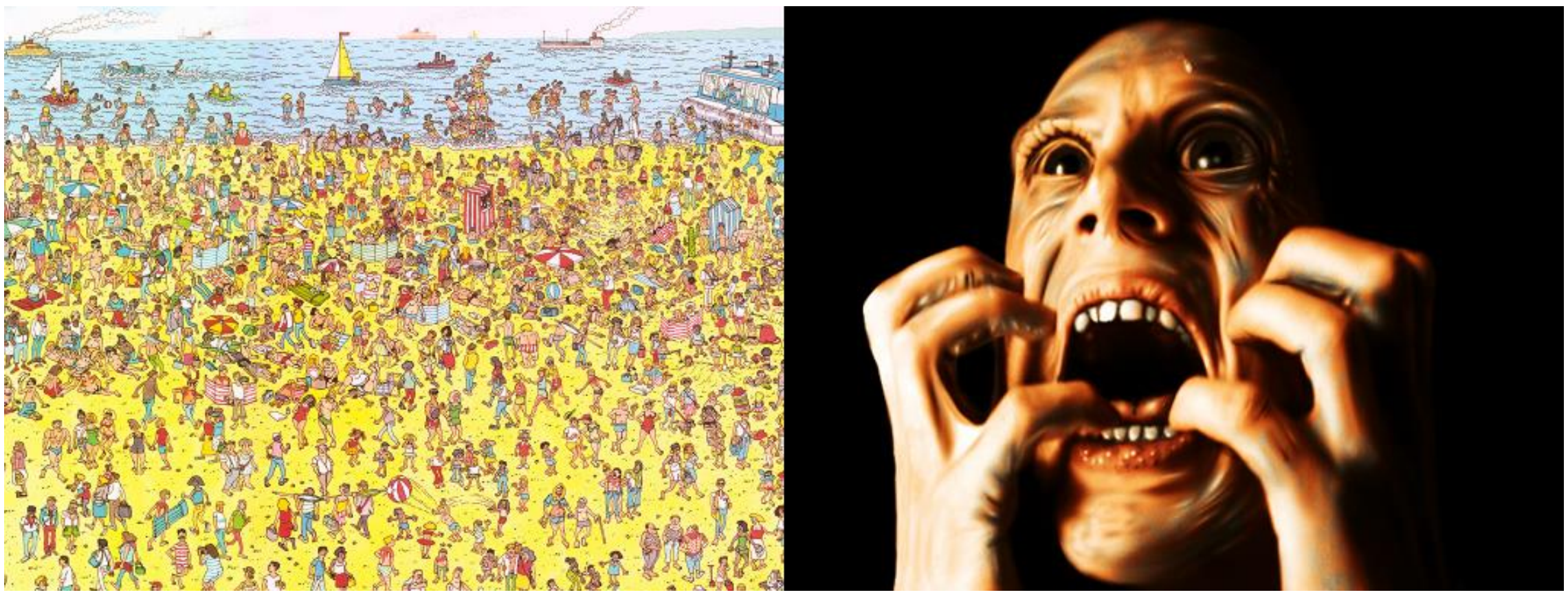

Figure 25: Images used for calibration process. On the left: Standard Waldo picture, with no Waldo in it. On the right: scary face to provoke a reaction on the player Arousal rating.

At the end of each condition, the participant was asked to fill the game experience questionnaire (GEQ) (IJsselsteijn et al. 2007). And once all the conditions had been completed, we then asked the participant to answer a final "biofeedback condition questionnaire" (BFCQ). Finally, the participants were thanked for their participation and escorted out of the lab. Given the low valence induced by all three conditions, as a reward, participants were given a small jawbreaker candy to raise their blood sugar levels.

## 5.3 Data Metrics

In order to answer the questions presented on the introduction of this chapter (Results and Discussion), the types of data sets were required. The first set is the various metrics derived from GEQ, being them: immersion, tension, competence, challenge, flow, positive affect and negative affect. The second cluster of data was retrieved from the BFCQ, with information on which condition the participant had more fun, if the participant perceived correctly the game mechanics present on each game condition, and if he noticed differences between each game condition. Finally, the third set of data is the participant's Arousal and Valence values throughout the three game conditions.

# 5.4 Results

## 5.4.1 Analysis of Variance

For assessing the statistical significance of the results, repeated measures analyses of variance (ANOVA) were conducted, using the different game conditions as the within-subject factor for each measurement. For this test we used all the metrics present on GEQ, the fun factor from the BFCQ and the Mean values for both Valence and Arousal from each of the 3 game conditions.

For GEQ components immersion ($\chi 2(2)$ =0.89459, $p > .05$), tension ($\chi 2(2)$ =0. 92432, $p > .05$), competence ($\chi 2(2)$ =0. 94236, $p > .05$), challenge ($\chi 2(2)$ =0. 9575, $p > .05$), flow ($\chi 2(2)$ =0. 90016, $p > .05$), positive affect ($\chi 2(2)$ =0. 80775, $p > .05$), negative affect ($\chi 2(2)$ =0. 86139, $p > .05$), and for the fun factor ($\chi 2(2)$ =0. 89974, $p > .05$), Mauchly's test showed that the assumption of sphericity had been met. For the Mean Arousal ($\chi 2(2)$ =0. 67718, $p < .05$) and Mean Valence ($\chi 2(2)$ =0. 5418, $p < .05$) it was violated. Therefore, degrees of freedom were corrected for both Mean Arousal and Valence using Greenhouse-Geisser estimates of sphericity, $\varepsilon = 0.75596$ and $\varepsilon = 0.68578$, respectively.

Statistical significance was unfortunately not achieved for the components: Competence: $F(2,44)= 0.6089$, $p > .05$), challenge: $F(2,44)= 0.3289$, $p > .05$, flow: $F(2,44)= 0.4879$, $p > .05$, fun: $F(2,44)= 3.1571$, $p > .05$, arousal mean: $F(1.5,33.26)=0.6694$, $p > .05$, and valence mean $F(1.37,30.17)=0.5974$, $p > .05$. The elements immersion: $F(2,44)= 12.115$, $p < .05$, tension: $F(2,44)=13.004$, $p < .05$, positive affect: $F(2,44)= 4.0187$, $p < .05$, and negative affect: $F(2,44)= 13.263$, $p < .05$ were all statistically significant. This is a sign that the different mechanics on each game condition significantly affected the player experience on some of the most important factors, such as, immersion, negative and positive affect.

Since the ANOVA tests revealed significant statistical differences between some components, Post-hoc analyses were conducted for them to identify which conditions differed from the others. The performed Tukey Post-hoc tests revealed a significant difference between the following conditions for each component: Immersion: N-BF and V-IBF ( (7.369565, 1.476), (8.091304, 0.9899096), $p < .05$), and between N-BF and NV-IBF ( (7.369565, 1.476), (8.66087, 1.067171), $p < .05$); Tension: N-BF and V-IBF ( (6.417391,1.435477), (7.543478,1.198418), $p < .05$) and between N-BF and NV-IBF ( (6.417391,1.435477), (7.678261,1.46999), $p < .05$); Positive affect: between N-BF and V-IBF ( (4.956522,2.499605), (5.8,2.704037), $p < .05$); and finally Negative affect: between N-BF and V-IBF ( (5.643478,2.366591), (4.676087,2.423365), $p < .05$). Overall the participants reported substantial differences between the V-IBF and N-BF conditions for all of the performed Post-hoc tests. Also, there were also noticeable alterations between NV-IBF and N-BF conditions in what regards the immersion and tension components.

### 5.4.2 Group Difference Analysis

Several independent two-tailed t-tests were conducted to compare the different metrics mentioned above (GEQ, BFCQ and Mean Valence/Arousal) between various: participant's gender, participant's gaming type and participant's affection toward horror game.

This first cluster of t-tests was conducted in order to assess differences between participant genders. There were significant differences in the scores for challenge (t(31.591) = 2.1672, p < .05), flow (t(41.498)= 3.4492, p< .05), negative affect (t(63.822)= -2.0347, p< .05) and fun (t(38.962)= 2.4372, p< .05). Unfortunately statistically significance was not achieved for the components: immersion (t(44.659) = 1.7074, p > .05), tension (t(39.841) = 0.1792, p > .05), competence (t(42.494) = 1.9717, p > .05), positive affect (t(53.071) = 1.9536, p > .05), arousal mean (t(44.252) = -0.135, p > .05) and valence mean (t(48.221) = -0.4295, p > .05). These results suggest that the game difficulty (challenge), but also the entertaining factor of the game (flow and fun) are distinctively perceived by both genders. Another curious but interesting result is that there was a statistical difference regarding negative affect, with male participants reporting a lower value than female participants ((M=4.503333, SD = 2.713778) and (M=5.612500, SD= 1.789902), respectively).

In the second wave of t-tests, we assessed whether there were statistically significant variances among gamer types (hardcore and soft players). Our tests revealed no discrepancies between player types on the following components: immersion (t(41.47) = 0.1287, p > .05), tension (t(45.988) = 0.6712, p > .05), competence (t(43.507) = -0.7166, p > .05), challenge (t(47.478) = -0.8262, p > .05), flow (t(45.391) = -0.9957, p > .05), positive affect (t(39.912) = -0.5614, p > .05), negative affect (t(33.446) = 1.448, p > .05), and fun (t(43.453) = 0.3074, p > .05). On the other hand, both mean arousal (t(41.415)= -4.0356, p< .05) and mean valence (t(25.89)= 2.2576, p< .05) revealed dissimilarities. These results seem to indicate that hardcore players have more neutral ratings of Mean Arousal (M=6.318667, SD = 0.818533) and Mean Valence (M=4.573333, SD = 1.28212) when compared with softcore players Mean Arousal (M=7.235238, SD = 0.7683558) and Mean Valence (M=3.897143, SD = 0.5857405). This suggests that hardcore players are not as susceptible to game *stimuli* as softcore players.

The third group of t-tests was conducted in order to evaluate if there were significant variations between the participants who liked and the ones who did not like this particular game genre. The components: immersion (t(48.336) = 1.5914, p > .05), tension (t(48.999) = -1.1088, p > .05), competence (t(47.726) = -0.2056, p > .05), challenge (t(41.849) = -1.3777, p > .05), flow (t(47.77) = 1.027, p > .05), negative affect (t(35.519) = -1.5187, p > .05), mean arousal (t(42.447) = 0.246, p > .05), and mean valence (t(47.679) = -1.2608, p > .05), did not achieve a noteworthy difference between the two groups. As of the elements: positive affect (t(44.94)= 2.38, p< .05) and fun (t(48.902)= 2.706, p< .05) both proven to be statistically significant. This supports the idea that if the participant enjoys this genre of game, he would have more fun and positive affect while playing it. And as we can take from the differences regarding fun, the people who liked this genre of game (M=3.916667, SD = 0.9286112) had a higher value of fun than the people who do not like this genre of game (M=3.185185, SD = 1.001423).

Finally we decided to conduct another set of tests to assess if female participants had more stable AV ratings than men, as Champion suggests on his work (Dekker and Champion 2007). Measured in terms of the standard deviation of the AV signal, the tests failed to achieve statistically difference on both Arousal ($t(39.525) = -0.4367$, $p > .05$) and Valence ($t(66.968) = 0.5332$, $p > .05$), which seems to present an interesting contrast to Champion's findings. We posit that perhaps this effect is not consistent among game genres (Champion's testbed game was a first-person shooter without any of the psychological terror aspects that VANISH presents). It is possible that the gamer type factor had an influence on this test, as they were unbalanced (75% of female participants were casual gamers). This suggests further study into the matter is necessary, but it is unfortunately not possible since Champion does not refer how gamer types were distributed across gender in his work.

### 5.4.3 Condition Variance Analysis

Despite the fact that no statistical significance was found on the ANOVA tests for mean AV ratings, a more detailed analysis showed that these ratings still demonstrate clear trends (see Figure 25). V-IBF elicits higher arousal values than the N-BF condition and NV-IBF even higher values than the previous two. The reason behind this last one might be due to the fact that NV-IBF has a strong focus on triggering mechanics in order to induce a significant variation on the player's emotional state. In terms of Valence, the N-BF condition presented the lowest valence ratings, with V-IBF having again higher values than N-BF, and NV-IBF following the trend and having the highest values. The fact that NV-IBF has again the highest values and closer to a neutral position, can be explained by its mechanics that, as mentioned in section 3.2.4, strive to maintain the player in a balanced emotional state. As of the V-IBF having higher values than N-BF can quite possibly be due to the system rules only covering character mechanics and not controlling event frequency and intensity.

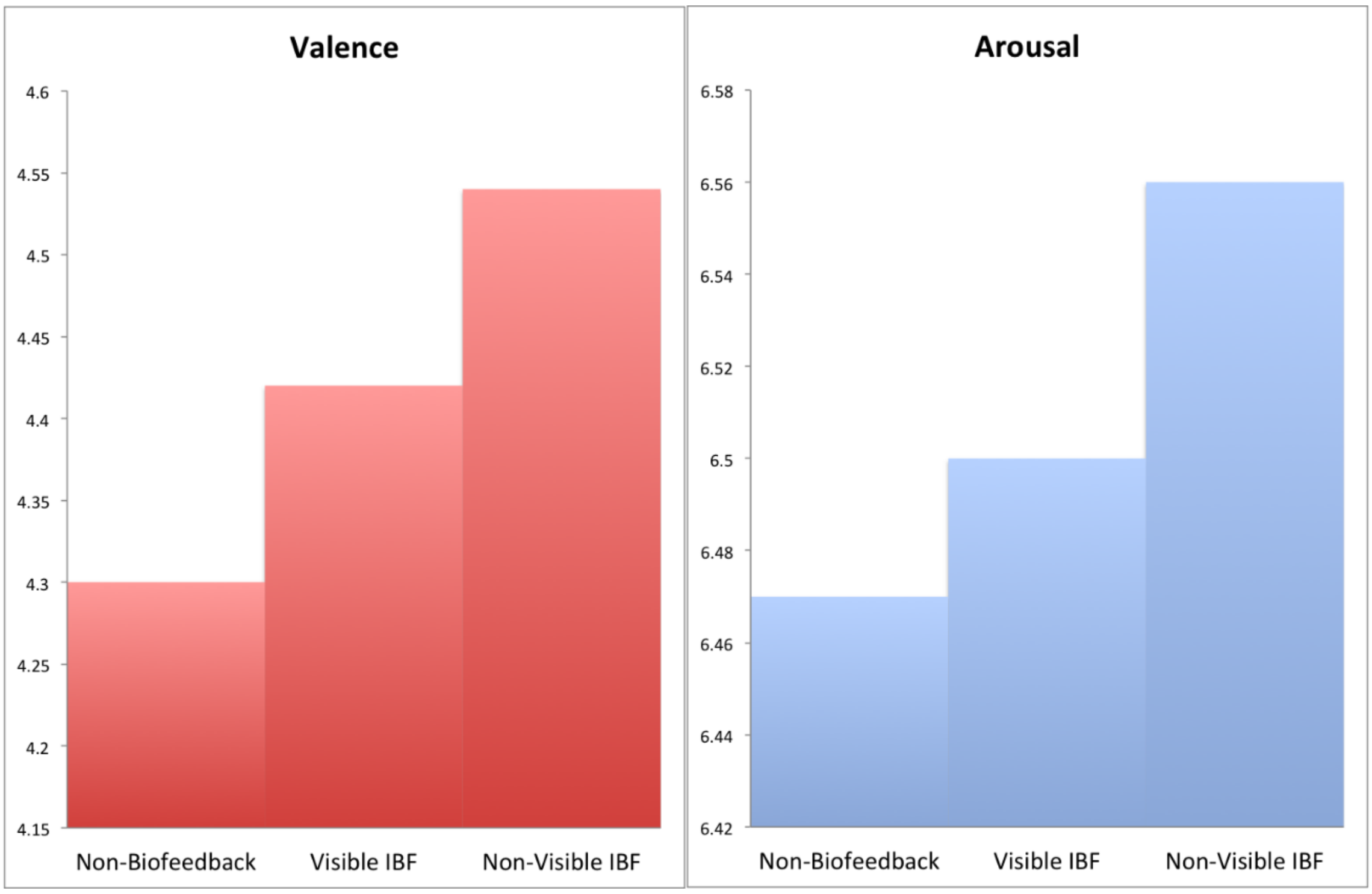

Figure 26: Mean Arousal and Valence on the three game conditions.

There were also notable results on the mean scores of GEQ. As we can see on Figure 26, there is a growth in immersion, tension and fun components on the conditions with a biofeedback system (V-IBF, NV-IBF). Another noteworthy outcome is the decrease of negative affect, with the NV-IBF condition reporting the lowest value, this can be explained by the mechanics used in this system (see 3.2.4);

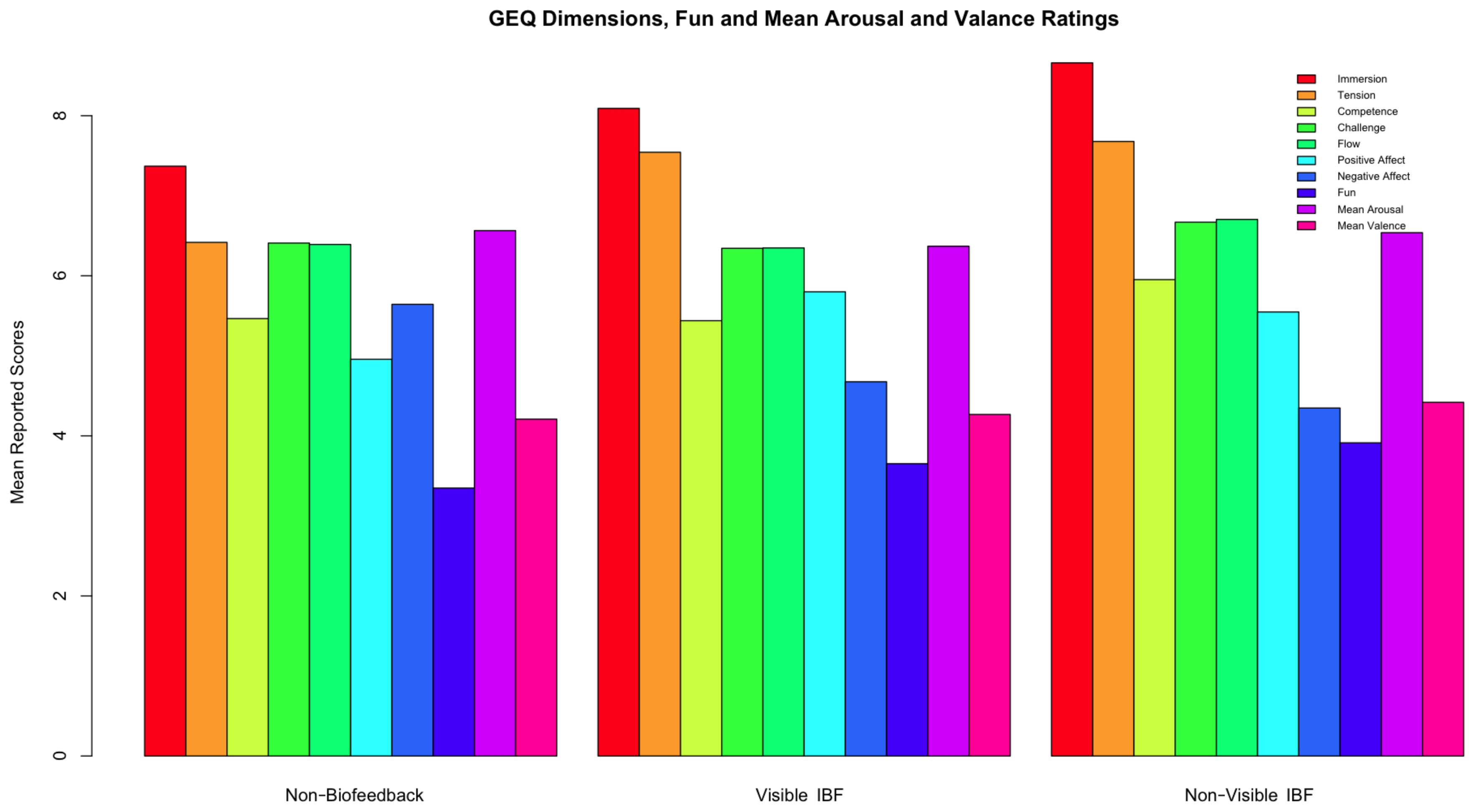

Figure 27: Mean scores for GEQ components in each game condition.

### 5.4.4 Condition Identification

When the participants were asked if they noticed differences between conditions, apart from one of them, all answered positively (see Table 8).

Table 8: Participants answers regarding condition identification.

| Noticed differences between | Condition Order | | |
|---|---|---|---|
| | Condition | Condition | Condition |
| Yes | V | NB | NV |
| No | NV | V | NB |
| Yes | NB | NV | V |
| Yes | NB | V | NV |
| Yes | V | NV | NB |
| Yes | NV | NB | V |
| Yes | V | NB | NV |
| Yes | NV | V | NB |
| Yes | NB | NV | V |
| Yes | NB | V | NV |
| Yes | V | NV | NB |
| Yes | NV | NB | V |
| Yes | V | NB | NV |
| Yes | NV | V | NB |
| Yes | NB | NV | V |
| Yes | NB | V | NV |
| Yes | V | NV | NB |
| Yes | NV | NB | V |
| Yes | V | NB | NV |
| Yes | NV | V | NB |
| Yes | NB | NV | V |
| Yes | NB | V | NV |
| Yes | V | NV | NB |

From the 23 subjects, 12 successfully identified every condition, while only 2 erred on the recognition of all conditions. The remaining 9 participants correctly identified one condition but failed the other two. From this last group, 6 mistook N-BF for V-IBF. The reason behind this might be related to the fact that most of the mechanics chosen for the V-IBF only being triggered when the player's emotional state deviates from its neutral values. Also, some of these mechanics might be slightly difficult to compare, since the player does not have a base value to compare to, i.e. - character run speed changing with player Arousal level. From the other 3 participants, only one mistook the NV-IBF and V-IBF conditions while the other two mistook NV-IBF for N-BF. Although we had some initial concerns that due to the modified mechanics on the NV-IBF condition being focused on procedural generation algorithms and artificial intelligence, the participants would not be able to successfully identify this condition without an habituation period. However these results proved that our extra attention when developing these mechanics revealed to be more efficient than the cautious point of view that led us to such concerns.

### 5.4.5 Condition Preference

When asked about their preference of game condition, participants' opinions strongly leaned towards the NV-IBF – which registered 43% of the total top preferences. The V-IBF registered in second place with 39% of participants' preferences and the NBF came in third place with only 13%. One participant reported being indifferent to each of the gameplay conditions, representing approximately 5% of the total participant population (see Figure 27). We find these to be very positive results, since they clearly suggest that the conditions with a biofeedback system had a greater impact on the participant affection towards the game. It is also very interesting the slightly inclination towards the NV-IBF when compared with V-IBF, which might suggest that even though the V-IBF approaches the player to the game avatar and gives the player some control over it, it does not provide the same advantages towards the game objectives as the NV-IBF condition.

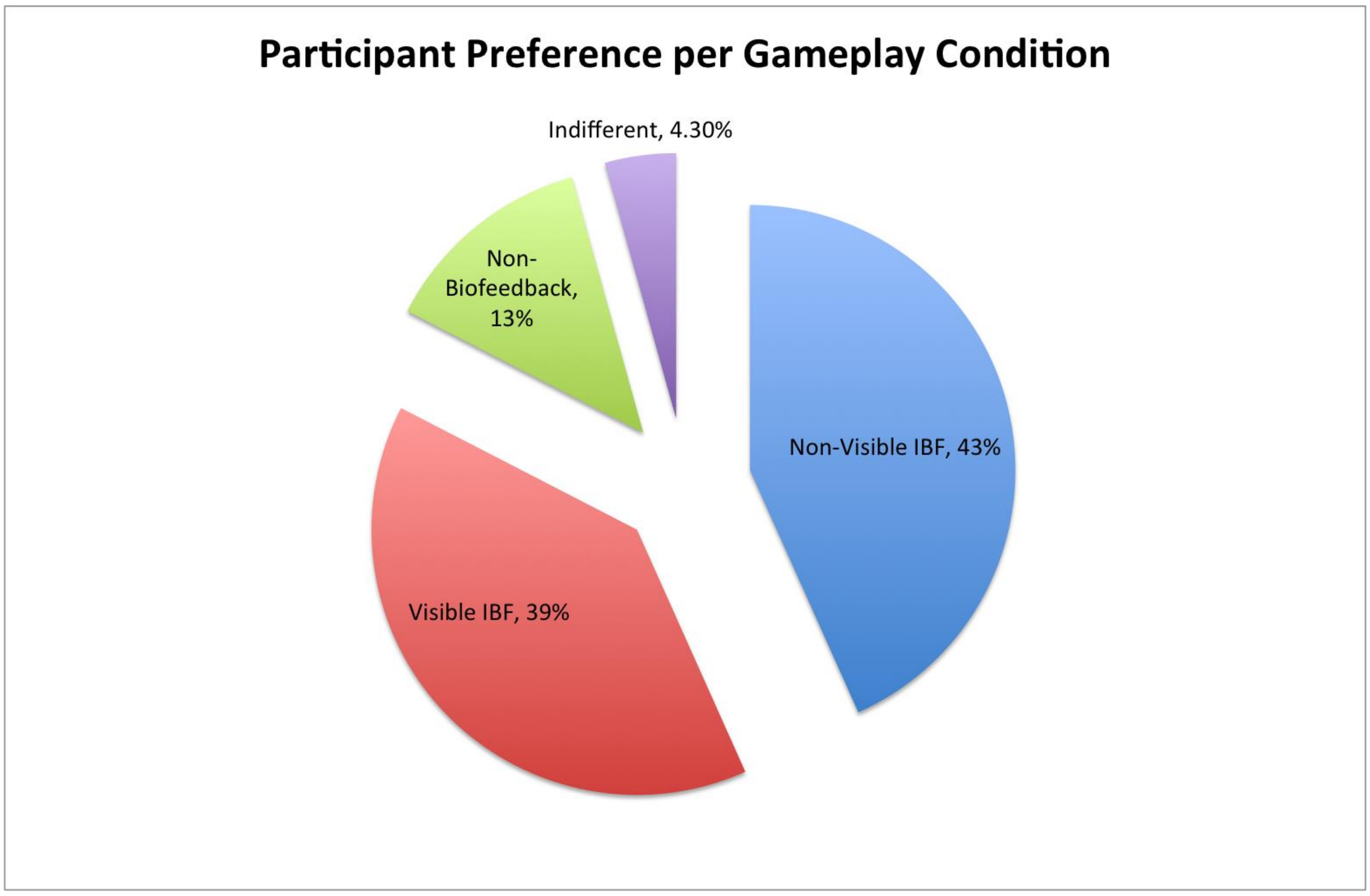

Figure 28: Participant preference per gameplay condition.

### 5.4.6 Participant Commentaries

When asked about the potential of this system, all of them answered positively, with a few noticing that some aspects had still to be ironed-out.

Apparently two participants felt like the game was being condescending or toying with them (P3) in the NV-IBF condition:

"*I found a dead end and when I turned back and took the same path, it was all different, for moments I thought it was me, but no, the map really changed*" (P17).

This was most probably due to the fact that they only noticed on that condition that the level generation was dynamic. A significant number of participants confirmed that the NV-IBF version was indeed the one where they felt more "immersed" (P6) into the game and it felt like it was "driven by a purpose" (P14) due to the timing chosen for the events to happen:

"*It totally caught be off guard, I was very agitated on the beginning of the game, but the game pacing became progressively slower, and when I felt more relaxed, the creature appeared and started chasing me!"*(P14).

Regarding the V-IBF condition, several participants stated that they felt like the game mechanics were intensifying their emotional state:

"*When the character started gasping and I could hear his heart beating faster and louder, it made me even more agitated*" (P10).

There was also gameplay/mechanism improvement suggestions by the participants. Some said that on the V-IBF condition the running mechanism was completely imperceptible:

"*I could not perceive if there was a difference on the running speed because I was always running and the character was continuously exhausted*" (P6).

He also offered a solution:

"*It would be easier if there was a stamina gauge on the user interface, so the player can have a better perception of how much stamina left the character has.*" (P6).

Still on the V-IBF condition, one participant that triggered the faint mechanism, said:

"*When the character fainted, for moments I thought I had lost the game, because I could not see or hear anything*" (P5).

He also proposed a good solution:

"*A ringing sound when the screen is black would probably hint the player of what is happening with the character*" (P5).

There was also the suggestion to make the intention of the evasion tunnels more evident:

"*I saw the little tunnels but I never had the curiosity of going in there, it was pretty dark and I did not know what was inside*" (P13).

Also after acquiring information of the purpose of these tunnels they said:

"*When I was running from the creature all I could do was look ahead and press the sprint key, if I knew I could hide from it on those tunnels I would probably search for them*" (P21).

Related to the creature AI, a few players said that it was "unfair":

"*When it started appearing it was always running from me, so I kept running into it, when suddenly it lunged at me! I found it really unfair since there was no previous intention of it*" (P19).

# Chapter 6

# Conclusions

## 6.1 Goals

The main objective of this work was to develop a robust and completely game-independent framework of an indirect biofeedback system. We also had to create more than one game adaptations and subsequently the implementation of their rules on the CLEARS sub-system, in order to retrieve good quality results. The results showed us that the game conditions that had our biofeedback framework working had a preference of 82%, which clearly shows that the $E^2$ system contributes significantly to the player affection towards the game.

While developing our main objective, we came across a challenge that originated the EET tool. Although initially we did not have strong intentions on advancing with this tool, when we studied the current State of the Art, we realized the potential that it had. With the results we got from it, we can consider that the development of this tool was a success.

In sum, we consider that our work was overall a success, and have exceeded our initial expectations, with the great contribution of two tools for the research and development of games with biofeedback systems.

## 6.2 Future Work

Since we have now two tools, our future work now has a distinct focus. Firstly we would like to finish our implementation of the $E^2$ architecture, with the conclusion of the $ARE^2S$ component. Even though we created its fundamental algorithms on the EET tool, we still have to implement it in real-time on the $E^2$ framework and have it communicate successfully with the other components.

The second focus should be to enhance the GLaDOS sub-system, and instead of having to develop it inside the game, we could create an easy API and have it communicate with the game. With this modification, the people in charge of the game would not have to understand how the biofeedback system works, and only had the task to learn how to communicate with it using the API.

A third focus would be to have the framework work dynamically by automatically constructing the player's Affective Reaction Profile (ARP) (Nogueira, Rodrigues, Oliveira and Nacke). In order to do this we would firstly need $ARE^2S$ to update the player's ARP profile – possibly using an exponential averaging function as suggested by (Nogueira, Rodrigues, Oliveira and Nacke) – and then building a statistical model that generalized the player's emotional response over the AV space. Secondly, we would need to implement Nogueira's ER version of CLEARS that dynamically infers which event has the highest probability of eliciting the desired emotional state on the player, given his current ARP.

The fourth and final focus, would be to augment the EET tool by increasing the amount of parameters it takes, with the possibility of specifying which signal data should be used for either Arousal or Valence, and their set of rules for the AV assessment. The tool output could also be improved, and instead of text we could display visual data, which can be better perceived by the user.

# Appendix A

## A.1 Game Experience Questionnaire

All dimensions were measured in a 5-point Likert scale. From the 7 dimensions, all of them are measured with 5 questions, with the only exception of immersion which is measured with 6. The following table presents the translated questions of GEQ (IJsselsteijn et al. 2007):

| DIMENSÃO | ITEM | QUESTÃO |
|---|---|---|
| Imersão | 3 | Estava interessado na narrativa do jogo |
| (Immersion) | 14 | O jogo era visualmente apelativo |
| | 20 | Senti-me imaginativo durante o jogo |
| | 21 | Senti que me era permitido explorar o mundo |
| | 30 | Achei o jogo impressionante |
| | 33 | Foi uma experiência de jogo rica/cativante |
| Flow | 5 | Senti-me completamente absorvido pelo jogo |
| (Flow) | 15 | Esqueci-me de tudo que me rodeava |
| | 28 | Perdi noção do tempo |
| | 31 | Estava profundamente concentrado no jogo |
| | 34 | Estava desligado do mundo exterior |
| Competência | 2 | Senti-me capaz de usar as minhas habilidades |
| (Competence) | 12 | Senti-me forte |
| | 17 | Senti que tive uma boa performance |
| | 19 | Senti-me bem sucedido com o meu desempenho |
| | 23 | Consegui atingir rapidamente os objectivos do jogo |
| Tensão | 7 | Senti-me tenso |
| (Tension) | 9 | Senti-me inquieto |
| | 24 | Senti-me incomodado (*annoyed*) |
| | 27 | Senti-me irritado |
| | 32 | Senti-me frustrado |
| Desafio | 8 | Senti que estava a aprender |
| (Challenge) | 13 | O jogo foi difícil |
| | 26 | Senti-me estimulado |
| | 29 | Senti-me desafiado |
| | 36 | Tive de me esforçar bastante |
| Afecto Positivo (Positive Affect) | 1 | Senti-me satisfeito |
| | 4 | Conseguia rir-me sobre os eventos que ocorriam |
| | 6 | Senti-me feliz |
| | 16 | O jogo foi agradável |
| | 22 | Gostei da experiência |
| Afecto Negativo (Negative Affect) | 10 | Pensei sobre outras coisas que não o jogo |
| | 11 | A experiência foi cansativa |
| | 18 | Senti-me aborrecido |
| | 25 | Fui distraído |
| | 35 | A história era aborrecida |

## A.2 Biofeedback Condition Questionnaire

**Quanto tempo despende semanalmente em video jogos? ***

- ○ 0-4 horas
- ○ 4-8 horas
- ○ 8-16 horas
- ○ Mais de 16 horas

**Gosta de jogos de terror? ***

- ○ Sim
- ○ Não

**Costuma jogar jogos de terror? ***

- ○ Sim
- ○ Não

**Quais as suas plataformas de jogo preferidas? ***

- ☐ PC
- ☐ PlayStation 2/3
- ☐ XBox
- ☐ Nintendo Wii
- ☐ Plataformas Móveis
- ☐ Dispositivos Retro

**Questão 1: Quão divertida foi cada condição? (1-aborrecida, 5-o mais que já me diverti neste tipo de jogo) ***

Condição #1:
Condição #2:
Condição #3:

**Questão 2: Notou alguma diferença entre alguma das condições? ***

- ○ Sim
- ○ Não

**Questão 2.1: Se sim, entre quais condições notou diferenças? ***

- ☐ Entre todas elas
- ☐ Entre a 1ª e a 2ª
- ☐ Entre a 1ª e a 3ª
- ☐ Entre a 2ª e a 3

**Questão 3: O entrevistador irá agora informa-lo/a sobre quais as diferenças e mecanismos de jogo alterados por cada condição de jogo. No entanto, não lhe dirá qual a ordem em que as jogou. Por favor tente identificar em qual condição observou cada mecânica em ação**

**Mecânica de jogo #1: Velocidade e duração do tempo de sprint. ***

Descrição de mecânica: Quando o jogador tem um valor de arousal mais alto, pode sprintar mais rápido (adrenalina elevada), mas a personagem cansa-se mais rapidamente. Inversamente, quando está calmo não corre tão rápido (velocidade normal), mas pode correr por mais tempo (resistência).

- ☐ 1ª condição
- ☐ 2ª condição
- ☐ 3ª condição
- ☐ Não observada

**Mecânica de jogo #2: Respiração e ritmo cardíaco da personagem. ***

Descrição da mecânica: A respiração e ritmo cardíaco da personagem são afectados pelo nível de arousal do jogador. Quanto maior, mais assustada a personagem fica e consequentemente, mais suspiros de medo e alto o seu ritmo cardíaco.

- ☐ 1ª condição
- ☐ 2ª condição

- [ ] 3ª condição
- [ ] Não observada

**Mecânica de jogo #3: Desmaiar. ***

Descrição da mecânica: Quando o jogador atinge o nível máximo (10) de arousal, a personagem começa a ficar tonta e desmaia.

- [ ] 1ª condição
- [ ] 2ª condição
- [ ] 3ª condição
- [ ] Não observada

**Mecânica de jogo #4: Sanidade mental. ***

Descrição da mecânica: À medida que a experiencia do jogador começa a ficar mais negativa (valence baixo), a probabilidade de a personagem começar a ficar maluca aumenta. Quando a sanidade dela é suficientemente baixa (i.e. valence muito baixo), esta começa a ter halucinações (vê bichos nas paredes, começa a ter visão em túnel, o ecrã escurece, etc.).

- [ ] 1ª condição
- [ ] 2ª condição
- [ ] 3ª condição
- [ ] Não observada

**Mecânica de jogo #5: Criatura ***

Descrição da mecânica: A probabilidade da criatura aparecer varia conforme o nível de agitação do jogador. Se o jogador tiver um nível de arousal alto a criatura aparece menos vezes, se o jogador tiver um nível de arousal mais baixo a probabilidade da criatura aparecer aumenta.

- [ ] 1ª condição
- [ ] 2ª condição
- [ ] 3ª condição
- [ ] Não observada

**Mecânica de jogo #6: Eventos ambiente ***

Descrição da mecânica: Semelhante à mecânica anterior, desta vez com eventos ambientes (canos explodem/caem, luzes explodem).

- [ ] 1ª condição
- [ ] 2ª condição
- [ ] 3ª condição
- [ ] Não observada

**Mecânica de jogo #7: Objectivos e Saida ***

Descrição da mecânica: A probabilidade de as salas de objectivos e a saída aparecerem varia com o nível de valence do jogador. Se o jogador estiver a sentir uma experiência negativa, o jogo aumenta a probabilidade de aparecerem estas salas. Se a experiência tiver a ser muito positiva, a probabilidade diminui para criar mais estimulo ao jogador.

- ☐ 1ª condição
- ☐ 2ª condição
- ☐ 3ª condição
- ☐ Não observada

**Mecânica de jogo #8: Tuneis ***

Descrição da mecânica: Quando a criatura se encontra a perseguir o jogador, a probabilidade de aparecer um túnel varia com o valence do jogador. Se o jogador estiver a sentir uma experiência negativa, a probabilidade de aparecer um túnel aumenta, no caso da experiência ser positiva, então a probabilidade diminui.

- ☐ 1ª condição
- ☐ 2ª condição
- ☐ 3ª condição
- ☐ Não observada